\newcommand{\be}{\begin{equation}}
\newcommand{\ee}{\end{equation}}
\newcommand{\bea}{\begin{eqnarray}}
\newcommand{\eea}{\end{eqnarray}}

\newcommand{\Prob}{{\rm Prob}}

\documentclass{arxiv}

\copyrightyear{2014}
\pubyear{2014}
\usepackage{graphicx}
\usepackage{threeparttable}
\usepackage{multirow}
\usepackage{hhline}

\begin{document}
\firstpage{1}

\title[Protein identification with accurate statistics]{Mass spectrometry based protein identification with accurate statistical significance assignment}
\author[Alves and Yu]{Gelio Alves and Yi-Kuo Yu\footnote{to whom correspondence should be addressed}}
\address{National Center for Biotechnology Information,\\
 National Library of Medicine, National Institutes of Health\\
 8600 Rockville Pike, Bethesda, MD 20894, USA
}

\history{Received on XXXXX; revised on XXXXX; accepted on XXXXX}

\editor{\vspace*{-0.3in} Associate Editor: XXXXXXX}

\maketitle

\begin{abstract}

\section{Motivation:}
Assigning statistical significance accurately  has become 
increasingly important as meta data of many types, often assembled in hierarchies, 
are constructed and combined for further biological analyses. Statistical inaccuracy 
 of meta data at any level may propagate to downstream analyses, 
 undermining the validity of scientific conclusions thus drawn.  
 From the perspective of mass spectrometry based proteomics, even though 
 accurate statistics for peptide identification can now be achieved,
  accurate protein level statistics remain challenging.
\section{Results:}
We have constructed a  protein ID method that combines 
 peptide evidences of a candidate protein based on a rigorous formula derived earlier; in this formula the database $P$-value of every peptide is weighted, prior to the final combination, according to the number of proteins it maps to. We have also shown that this protein ID method provides accurate protein level 
 $E$-value, eliminating the need of using empirical post-processing 
 methods for type-I error control. Using a known protein mixture, we find that this protein ID method, when combined with the Sori\'c formula,
  yields accurate values for the proportion of false discoveries. 
In terms of retrieval efficacy, the results from our method are
 comparable with other methods tested. 
\section{Availability:}
The source code, implemented in C++ on a linux system, is available for download at
{\tt ftp://ftp.ncbi.nlm.nih.gov}\\{\tt /pub/qmbp/qmbp\_ms/RAId/RAId\_Linux\_64Bit}

\section{Contact:} \href{yyu@ncbi.nlm.nih.gov}{yyu@ncbi.nlm.nih.gov}
\vspace*{-0.25in}
\end{abstract}

\section{Introduction}
 Peptide identifications (ID) via mass spectrometry (MS) have become the 
  central component in modern proteomics; this component, combined with additional 
   analyses, routinely yields pragmatic meta data, including  
 protein ID, protein quantification, protein structure 
 and protein associations~\citep{ZFSBetal_2013}.   
 These meta data, especially the associated statistical significance assignments, 
 need to be as accurate as possible because they 
 often form the building blocks for investigations at the systems biology level
  and influence the scientific conclusions drawn henceforth. 
 In this paper, we focus on protein ID, in particular on improving the
 accuracy of statistical significance assigned to proteins identified.  
 
 The need for robust developments towards accurate statistical significance 
 assignments has been advocated~\citep{NM_2012,HWYH_2012} despite the
  existence of many  protein ID methods~\citep{MA_2008,SN_2012,LR_2012}. 
 It has also been suggested~\citep{SSSGetal_2011} that 
 the primary cause of unreliable significance assignment for protein ID 
 can be attributed to inaccurate significance assignment for peptide ID.
 Frequently used error-control/significance-assigning methods for peptide 
 ID largely fall into two groups: proportion of false discovery (PFD),
  which is often incorrectly termed as false discovery rate~\citep{BH_1995}, 
  and spectrum-specific $P$-value/$E$-value~\citep{FB_2003,AOY_2007,PKKMetal_2008}. 
 Methods belonging to the first group, controlling type-I error globally only, 
 do not discriminate among identified peptides~\citep{EG_2007}. Methods belonging 
 to the second group, capable of assigning per-spectrum per-peptide significance,  
  can {\it properly} prioritize identified peptides 
  when reported $P$-/$E$-values are accurate; but the
   needed statistical accuracy is often unattainable due to improper heuristics or unjustifiable distribution assumptions~\citep{AOY_2007,SEGAL_2008,SSSGetal_2011}.

 Given a tandem MS (MS/MS) spectrum and  
 a quality score cutoff  ${\sf S}_c$,   
 the $E$-value $E({\sf S}_c)$ should reflect 
 the expected number of random peptides with scores the same as 
 or better than ${\sf S}_c$.
  (Similarly, the $P$-value $P({\sf S}_c)$ reflects the probability of finding a random 
  peptide with quality score ${\sf S} \ge {\sf S}_c$.) In general, the $E$-value
  is obtained by multiplying the $P$-value by the total number of {\it qualified 
  peptides} (whose masses fall in the range $[m_p-\delta,m_p+\delta]$ 
   with $m_p$ being the precursor ion's mass and $\delta$ the specified tolerance) 
   in the database searched. Thus, besides providing the user with the numbers 
   of false positives to anticipate, accurate $E$-value assignments   
   enable ranking of candidate peptides across different spectra and experiments. 
  In database searches in proteomics, the goal of accurate statistics 
  can be approached in at least two ways. 
  First, one may devise a scoring function whose 
  resulting score distribution can be analytically characterized 
  and thus used to infer the statistical significance~\citep{AOY_2007}; 
  if this is done correctly, the theoretical score distribution 
  should fit well the bulk part of the normalized score histogram obtained from
 scoring all {\it qualified peptides} in the database of interest.  
   Second, one may infer the spectrum-specific $P$-value via the normalized score 
   histogram obtained from scoring all possible peptides (APP)~\citep{AY_2008};
    in this case, the database dependence appears only in the $E$-value, which is 
    the $P$-value multiplied by the number of qualified peptides associated with
     the specified precursor ion mass and mass error tolerance.   
 Either way yields database-specific $E$-values. 
 Once a peptide $E$-value is obtained, one may transform it into 
  the peptide \underline{d}atabase \underline{$P$}-\underline{v}alue (DPV)~\citep{YGASetal_2006,AWWSetal_2008},
   representing the likelihood of obtaining,
   in the database chosen, at least one peptide scoring equal to or better than
    the prescribed threshold. When combining $P$-values of peptides associated 
    with a candidate protein, we use the peptides' DPVs.  
%
%
%
%

 Specifically, our proposed protein ID method combines 
 peptide evidences of a candidate protein using a rigorous formula derived earlier~\citep{AY_2011}; 
  in this formula the DPV of every peptide is weighted, prior to the final 
  combination, according to the number of proteins it maps to.  
Among the existing protein ID methods, the approach taken by~\citet{SSSGetal_2011}
 is closest to ours; both methods combine peptides' 
  spectrum-specific $P$-values.  
 There are, however, major differences between our method and that of~\citet{SSSGetal_2011}.
  First, in our method, each candidate peptide of a query spectrum receives a
 DPV, allowing multiple matching peptides per spectrum.  
 This is to take into account the possibility of peptide 
 co-elution~\citep{AOKWetal_2008}.
 For the method of~\citet{SSSGetal_2011}, only the best peptide match 
 per spectrum is considered
  and the peptide DPV thus represents the probability of 
 having the {\it best} match score no worse than the prescribed threshold
  when searching a database.  Since each random protein database 
  only contributes one best match score,
  searching {\it many} random protein databases is required for the $P$-value assignment.
  Second, the candidate peptides' $P$-values are combined differently.
 Our method, down-weighting contributions of 
  peptides mappable to multiple proteins, combines peptide DPVs 
  directly using a rigorous formula~\citep{AY_2011};  
 the method of~\citet{SSSGetal_2011} first transforms, for every candidate protein, 
 the $P$-values of its associated peptides  into $Z$-scores, combines them using 
  Stouffer's formula~\citep{Whitlock_2005}, and then transforms the combined $Z$-score 
  back to a final $P$-value with multiple hypotheses testing correction. 
  Third, the cutoff conditions for peptides' $P$-values are different. 
  Our method approximates DPVs~\citep{YGASetal_2006,AWWSetal_2008} by $E$-values, valid for small $E$-values, and retains all peptides whose $E$-values are less than one.  That is, we have a global cutoff condition.  
   For the method of~\citet{SSSGetal_2011}, the peptide cutoff $P$-value varies by candidate protein:  
   given a candidate protein, its corresponding peptides' $Z$-scores are first 
    sorted in descending order; the $k$th $Z$-score is chosen as the cutoff provided 
    that the maximum combined $Z$-score is reached while combining 
    the top $k$ $Z$-scores using the Stouffer's formula.

 There exist many other protein ID methods, for example, ProFound~\citep{ZC_2000}, ProteinProphet~\citep{NKKA_2003}, DBParser~\citep{YDDMetal_2004}, EBP~\citep{PLWAetal_2007}, PANORAMICS~\citep{FNC_2007}, PROVALT~\citep{MA_2008}, X!Tandem~\citep{FEB_2010}, Scaffold~\citep{Searle_2010} and npCI~\citep{SPSS_2013}, to name just a few.
  We refer the readers to recent review papers~\citep{HWYH_2012,SN_2012} for details and more comprehensive 
  listings of these methods. Although some of them do start with spectrum-specific 
 peptide $P$-values, they often {\it assume} certain parametric forms for
 the peptide score distributions when searching a random database; 
  other methods, however, only process outputs of specific peptide identification 
  tools, limiting their uses to certain platforms. 
  By discarding all but the {\it best} few peptide scores 
  per spectrum per database search,  the method of~\citet{SSSGetal_2011} 
  does not rely on the accuracy of the full peptide score 
  distribution from searching a random database and in principle can accept
   input from various peptide identification tools.   
  Our method is free from the aforementioned problems for different reasons. 
 Founded on a derived analytical formula,   
   our method can be applied in general and will yield accurate protein $P$-values 
     if the input  peptide DPVs (or $E$-values) are accurate. 
   When using peptide $E$-values reported by RAId\_DbS, even though 
  the parameters of the score distribution  
   are determined via maximum-likelihood, the functional form of  
   the score distribution is analytically derived~\citep{AOY_2007} 
   rather than assumed. When the statistical significances are obtained from
   RAId\_aPS~\citep{AOY_2010}, for every scoring function implemented, 
     the $P$-values are inferred by scoring APP instead of assuming that the score 
      histogram follows a specific form; the peptide
     $E$-values are then obtained via multiplying the $P$-values by the respective 
     numbers of qualified peptides.

%

The paper is organized as follows. 
 The mathematical underpinnings of our formalism will be described in the methods section. In the results section,  comparisons of our method with other approaches 
will be made; the accuracy of the reported protein $P$-value will be illustrated. Some technical but important issues will be addressed in the discussion section. 
To keep the paper focused, we relegate to supplementary information
 figures and tables that complement or corroborate the information contained  
 in the main text.  

\begin{methods}
\vspace*{-0.2in} 
\section{Methods}
\subsection{Statistical Protocols}
Weighting the contribution of each peptide in protein identification is important.
 It helps mitigate the issue of peptide degeneracy, where an identified
  peptide is a subsequence of multiple database proteins. 
  The optimal weighting scheme, however, can depend on the protein ID methodology 
  employed. For the purpose of our study, namely, devising a method that yields 
  accurate protein $P$-values, we opt for a simple weighting scheme: 
   a peptide's weight is inversely proportional to the number of 
   database proteins it maps to. Within a sample, when multiple spectral searches 
   identify the same peptide but with different significance levels, 
     only the most significant assignment of that peptide is 
    retained for further analyses. 

The foundation of our method is built upon a rigorous formula~\citep{Mathai_1983,AY_2011} 
that enables weighted combination of $P$-values. When the weights are all 
identical, this formula reduces to Fisher's 
formula~\citep{FISHER_1932A,Bharucha-Reid_1960}; when the weights are all
 different, this formula reduces to the formula of Good~\citep{Good_1955}. 
 A detailed derivation and generalization to incorporating nearly identical
  weights can be found in~\citep{AY_2011}, whose notation will be used to 
  briefly summarize the content of the formula.    

 Let us assume that a given protein contains $L$ identified peptides with 
  $P$-values. Let us further group these $L$ peptides, 
  according to the number of database proteins a peptide maps to,  
 into $m$ groups with $1\le m \le L$. 
 Within each group $k$, the $n_k$ peptide $P$-values are weighted equally;
  while peptide $P$-values in different groups are weighted differently. 
  

The weighting enters our formalism through the following quantities of interest
 \bea 
\tau &\equiv & \prod_{k=1}^m \left[ \prod_{j=1}^{n_k} p_{k;j} \right]^{w_k}  \; ,\label{Unify_tau} \\
Q &\equiv &  \prod_{k=1}^m \left[ \prod_{j=1}^{n_k} x_{k;j} \right]^{w_k} \; ,\label{Unify_q}
 \eea   
where each $p_{k;j}$ represents a reported peptide $P$-value, 
each $x_{k;j}$ represents a random variable drawn from an uniform, independent distribution over $(0,1]$ and each $w_k$ is a positive weight.   
 The quantity of interest $\Prob(Q \le \tau)$, representing the protein $P$-value, 
 was obtained earlier~\citep{AY_2011} and is repeated below for clarity. 
 
 Let $F(\tau) \equiv \Prob (Q \le \tau)$, one may show that 
\bea 
\hspace*{-10pt} F(\tau) &=& \left[ \prod_{l=1}^m  r_l^{n_l} \right] \sum_{k=1}^m \,
\sum_{\mathcal{G}(k)} 
\;\biggr\{
\frac{1}{r_k^{g_k+1}}  H(- r_k \ln \tau \, ,\, g_k) \times  \nonumber \\
&&  \hspace*{-25pt} \times 
 \biggr(  \prod_{j=1,j\ne k}^m  \frac{ (n_j-1+g_j)!}{(n_j-1)!  g_j!}  \frac{(-1)^{g_j}}{(r_j-r_k)^{n_j+g_j}}  \biggr) \biggr\} 
,
\label{F.unify.1}
\eea
where $r_k \equiv 1/w_k$ is the number of proteins a  group-$k$ peptide maps to, 
 $\sum_{\mathcal{G}(k)}$ enumerates each set of
 nonnegative integers $\{g_1,g_2,\ldots,g_m\}$ that satisfies  
 the $k$-dependent constraint $\sum_{i=1}^m g_i = n_k-1$, and 
the function $H$ is defined as 
\be \label{H.def}
H(x,n) \, \equiv \;  e^{-x}\, \sum_{k=0}^{n} \frac{x^k}{k!}  \; .
\ee
See the supplementary information for 
an example application of formula~(\ref{F.unify.1}). 

When searching a database with a prescribed peptide mass error tolerance $\delta$, 
 one often needs to score different numbers of database peptides 
 for spectra with different precursor ion masses. That is, the number of tested
  hypotheses (database peptides in the mass range $[m_p-\delta,m_p+\delta]$) 
  varies by the precursor ion mass $m_p$. The effect of 
  varying number of multiple hypotheses tested can be properly accounted for 
  by using the peptide DPVs~\citep{YGASetal_2006,AWWSetal_2008}
   for $P$-values ($p_{k;j}$) in eq.~(\ref{Unify_tau}); 
 given a quality score cutoff ${\sf S}_c$, the peptide DPV 
 is defined as
\be \label{E2P}
P_{\rm db}({\sf S}_c) = 1 - e^{-E({\sf S}_c)}\; , 
\ee   
 where $E({\sf S}_c)$ represents the expected
 number of peptides having score ${\sf S} \ge {\sf S}_c$,
  and the DPV $P_{\rm db}({\sf S}_c)$ 
 represents the probability of seeing one or more 
 peptides in a given random database with quality scores 
 ${\sf S} \ge {\sf S}_c$. Another advantage of using DPV is 
 that as a function of the quality score ${\sf S}$,  
  the $E$-value $E({\sf S})$, determined by the  search score histogram
  per spectrum and the number of qualified peptides (database-dependent), 
  correctly takes into account both the spectrum-specificity 
  and the database-specificity of scoring statistics. 

Since the $E$-value  
 specifies the expected number of random database peptides  
  having scores equal to or better than the given cut-off, a peptide
  with $E$-value larger than one is more likely to be a false positive 
   than a true positive. For this reason, when constructing the evidence peptide 
   set for identification of a protein, 
   we only include peptides with $E$-values less than one. This implies that only
  peptide DPVs less than $(e-1)/e$ are considered, leading to
   a combination of truncated $P$-values. Unfortunately, 
  combining truncated $P$-values, even though doable, is far more complicated 
  than using eq.~(\ref{F.unify.1}). However, two observations 
 simplify the matter. First, it is evident from  
    eq.~(\ref{E2P}) that the DPV approaches the $E$-value
    when the $E$-value is small.   Second, we note that 
    confidently identified proteins must contain evidence peptides with high 
    identification confidences (or small $E$-values). Therefore, for practical 
   uses, we may approximate the DPV by its corresponding $E$-value.
 Because only $E$-values less than one are considered, 
 the approximated DPVs (or simply the $E$-values) now encompass 
 the full range between zero and one. 
   Consequently, it is unnecessary to combine truncated $P$-values, and the simple  
   formula~(\ref{F.unify.1}) becomes applicable. 
   The protein $E$-value is then obtained via
   multiplying the protein $P$-value by a Bonferroni correction factor; in this case, the
   Bonferroni factor is the number of protein clusters (described below)  
    each having at least one evidence peptide with
   $E$-value less than one.   

We denote by a protein cluster 
a group of {\it entangled} proteins 
that share a substantial portion of evidence peptides. 
 To avoid exaggerating the number 
 of identified proteins, several existing methods~\citep{HWYH_2012} 
 report those entangled proteins as one. 
 Adopting the same idea, we implemented this strategy 
 via a transitive approach described below. 
  One first sorts the identified proteins by the number of identified 
   evidence peptides in descending order and using the
   rank of a protein in the sorted list as that protein's cluster index.  
  Starting with the first protein as the 
  reference protein, all other lower-ranking proteins sharing 
  at least $95\%$ of evidence peptides 
   with the first protein will have their cluster indexes changed to 
   that of the reference protein. One then moves the reference point (from the first) to 
   the second protein, all other lower-ranking proteins sharing at least $95\%$ of
    evidence peptides with the reference protein will have their cluster indexes 
    changed to that of the reference protein. The reference point is then moved to the 
    third protein and the process continues till the reference point moves through
    all proteins in the list. 
    The protein with most significant $P$-value within a cluster (containing
     one or more proteins) is called the head of that cluster, 
     the other proteins members of that cluster. 
     An exception to the aforementioned clustering rule, however, is 
    introduced to appropriately emphasize a protein's  
   evidence peptides that are not shared by other proteins. 
   We call evidence peptides of this kind {\it unique} peptides to a protein.
   When a protein has a unique evidence peptide with $E$-value less than $10^{-4}$,
    our method doesn't allow this protein to be a member protein of any cluster.

%

\subsection{MS/MS Datasets}
Sixty-three spectral datasets were categorized into four data groups. 
See supplementary tables (Table S1 to S4) for details.
Protein mixtures giving rise to spectral datasets 
were reduced with iodoacetamide, resulting in
 the addition of the carbamidomethyl group (57.07 Da) to cystine residues.  
 Each protein mixture was further digested with trypsin. 
 Among these spectral datasets, there are also dataset-specific parameters 
 such as the target database, the maximum number of missed cleavage 
 sites allowed, the precursor-ion mass error tolerance 
 and the product-ion mass error tolerance. 
  The dataset-specific parameters are given in the figure caption 
  to provide more information underlying the generation of the figures.

 For brevity, we shall denote the MS/MS spectra obtained from  
 a sample by SN followed by its sample index. For example, SN1 denotes the collection of
  MS/MS spectra acquired from mixture sample one.  
The first data group, SN1 through SN15, 
  contained MS/MS spectra from replicates of different dilutions of Sigma49, 
 a protein standard mixture composed of 49 know human proteins. 
 The second data group, SN16 through SN26, was downloaded from the Pacific 
 Northwest National Laboratory and contained spectra 
 from eleven whole-cell-lysate samples 
 of protein mixtures of {\it Escherichia coli K-12}. 
  The third data group, SN27 through SN30, consisted of spectra from four in-house 
 whole-cell-lysate samples of protein mixtures of {\it Escherichia coli K-12}. 
 Downloaded from PeptideAtlas database, the fourth data group (SN31 through SN63)  
 was composed of spectra from SDS-PAGE protein fractionation extractions of human 
 lung cells.    
 
\subsection{Protein Databases and Random Databases} 
 Because protein mixtures from {\it Escherichia coli K-12} and {\it Homo sapiens}  
 were analyzed using their corresponding MS/MS spectra,  
 protein databases for both organisms were thus required. 
 From UniProt \href{http://www.uniprot.org/downloads}{http://www.uniprot.org/downloads}, 
 we downloaded 4,303 non-redundant protein sequences of {\it Escherichia coli K-12}. 
 A non-redundant {\it Homo sapiens} protein database, containing 
 $31,236$ protein sequences, was obtained from the NCBI site \href{ftp://ftp.ncbi.nlm.nih.gov/refseq/H_sapiens/mRNA_Prot/}{ftp://ftp.ncbi.nlm.nih.gov/refseq/H\_sapiens/mRNA\_Prot/}. 

When analyzing statistical significance, it is often required to have random (decoy) 
databases in addition to the organismal (target) databases. 
One common problem when using random databases is that for 
a given precursor ion mass 
the numbers of qualified peptides in the random database and in the organismal database may significantly differ. This causes an additional uncertainty in  assessing statistical significance~\citep{EG_2007,WWZMetal_2009}.
We can avoid this problem by ensuring that the numbers of qualified peptides per spectrum are identical for both the random and the organismal databases: 
 for each qualified peptide in the organismal database, we generate a 
 corresponding random peptide by randomly shuffling its amino acids.  

\end{methods}

\section{Results}
The results will be described in the following order. 
First, we illustrate that our $E$-value
 assignments are accurate at both the peptide and the protein levels. 
 We further show that using the formula proposed by \citet{Soric_1989}, 
 our reported PFDs agree well with the target-decoy PFDs. 
Second,  our protein $E$-value accuracy is compared 
with that of using the formulas in~\citep{SSSGetal_2011}. 
 By extending the formula of Sori\'c for the method of~\citet{SSSGetal_2011}, 
 we also evaluate the agreement between their reported PFDs 
 and the target-decoy PFDs. 
Benchmarking with some of the existing protein ID 
  methods will be described in the third part. \vspace*{-0.1in}
\subsection{$E$-value accuracy}
\label{sec:E-value}
\begin{figure}[htbp]
\begin{center}
\vspace*{-0.1in}
\includegraphics[width=1.0\columnwidth,angle=0]{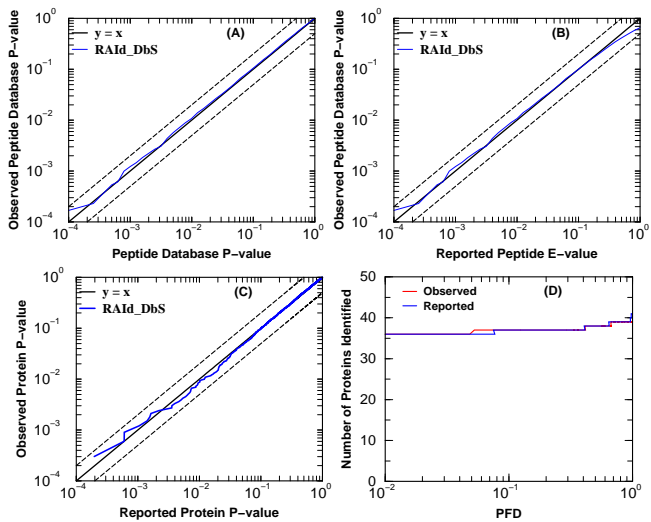}
\vspace*{-0.3in}
\end{center}
\caption{ {\bf Assessment of $E$-value accuracy}. 
In panels A, B, and C, the closer the displayed curves are to the $y=x$ line
 the better. In panel D, the closer the two displayed curves are 
 to each other the better. See section~\ref{sec:E-value} for more details.  
 For panels A, B, and C, 
spectral dataset SN 26 ({\it Escherichia coli K-12} whole cell lysate)
 is used to search the {\it Escherichia coli} database
 with  mass accuracy $\pm$ 0.033 Da. for precursor and product ions.
 For panel D, spectral datasets SN 13-15 (Sigma49 protein standard mixture) 
 is used to search the {\it Homo sapiens} 
 database with  precursor ion accuracy $\pm$ 0.033 Da. and product ion accuracy $\pm$ 0.8 Da. \vspace*{-0.1in}} 
\label{Fig:1}
\end{figure}

 The input peptide DPVs for our protein ID method
  are obtained via eq.~(\ref{E2P}) using the $E$-values reported by
  RAId\_DbS. For this reason, 
  the input peptide DPVs (for protein 
  ID) are synonymous with the reported peptide DPVs 
  (from RAId\_DbS). As mentioned earlier, the statistical accuracy of
 our protein ID method relies on the DPVs 
 for the evidence peptides being accurate.  
 We therefore start by comparing the input peptide DPV with 
  its definition. In panel A of Figure~\ref{Fig:1}, 
  the abscissa records the peptide DPV, while the ordinate
   displays the {\it observed} DPV ({\it i.e.}, 
   fraction of spectra having at least one or more matching peptides 
   with reported DPVs smaller than the specified threshold).  
 The agreement between the observed DPV and the reported 
 DPV 
 indicates that the peptide
  DPVs used as input for our protein ID method are accurate. 

To assess whether approximating peptide DPVs by
 their corresponding $E$-values for $E$-values less than one 
 is reasonable or not, we plot in panel B of Figure~\ref{Fig:1} the observed 
 peptide DPVs versus $E$-values. As expected, when $E$-values are close to one, there is certain degree of disagreement; while for small
  $E$-values, the agreement is excellent. To assess the accuracy of the
   protein $P$-values reported by our eq.~(\ref{F.unify.1}), 
   we compare them with the observed protein $P$-values. 
   As described in the method section, the reported proteins appear in clusters, each represented by a {\it head} protein and its $P$-value. 
 The observed protein $P$-value is defined as the 
   fraction of identified protein clusters (whose member proteins each containing at least one evidence peptide with $E$-value less than one) that have reported $P$-values smaller than a given threshold. As shown in panel 
   C of Figure~\ref{Fig:1}, good agreement between the reported protein 
   $P$-value and the observed protein $P$-value is obtained, 
   indicating that our reported protein $P$-values are accurate. 
   More protein $P$-value accuracy assessment examples can be found in
   Supplementary Figures S1, S2, and S3.  
   With an accurate protein $P$-value, one can also obtain 
  its corresponding protein $E$-value by multiplying it by the total number 
  of protein clusters. In Supplementary Figure S4, we show that 
  reported protein $E$-values obtained this way are accurate. 
  
By having accurate protein $E$-values, one can avoid the uncertainty 
 associated with using a decoy database~\citep{GBKP_2011} 
 while estimating the proportion of false discoveries. In panel D of Figure~\ref{Fig:1}, we plot two PFD curves:
 one is computed using the reported protein $E$-value to estimate the number of 
  false identifications (hence the PFD), while the other is computed using the 
  observed PFD obtained from known target protein content in the sample (Sigma49).
   The excellent agreement between the observed PFD and the reported PFD 
   indicates that one should be
   able to trust the PFD estimated from accurate reported protein $E$-values.  
  More accuracy assessment examples of the reported PFD can be found in 
  Supplementary Figure~S5.  \vspace*{-0.1in}   
    
\subsection{Comparison with an EVD-based method}
\label{sec:EVD-comp}
Since the method of~\citet{SSSGetal_2011} is closest to ours, we also implemented 
their method and compute equivalent quantities for comparison.
Following the Supplementary Material of~\citep{SSSGetal_2011}, 
we have implemented 100 random databases each containing 10,000 random amino acid sequences. 
 However, instead of generating sequences of uneven length, we opt for uniform length 
 (each sequence is of length 350) and generate these random sequences using the 
 background amino acid frequencies of ~\citet{RR_1991}. 
 The EVD parameters are obtained by using only the best score per database search 
 and by applying standard procedures described in~\citep{SSSGetal_2011}. The effect of database size difference, leading to rescaling 
 of the $\alpha$ parameter, is done the same way as in~\citep{SSSGetal_2011}.  

\begin{figure*}[htbp]
\begin{center}
\vspace*{-0.1in}
\includegraphics[width=0.9\textwidth,angle=0]{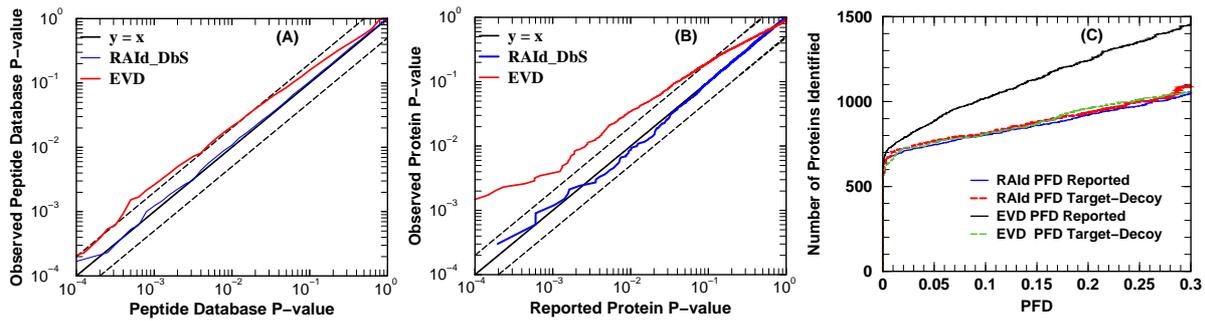}
\vspace*{-0.2in}
\end{center}
\caption{ {\bf Statistical Accuracy Comparison}. 
 Except that results from two methods are being displayed,  
 panels A, B, and C display similar information respectively to panels A, C and D of Fig.~\ref{Fig:1}. See text in section~\ref{sec:EVD-comp} for more details.
 \vspace*{-0.1in}} 
\label{Fig:2}
\end{figure*}

A moment of reflection reveals that the best match $P$-value of~\citep{SSSGetal_2011} is in fact the DPV~\citep{YGASetal_2006,AWWSetal_2008}.
We therefore plot in panel A of Figure~\ref{Fig:2} the reported 
peptide DPVs against the observed peptide DPVs.
The result indicates that the peptide DPV reported by~\citet{SSSGetal_2011} is quite accurate, with an uncertainty of a factor of $2$ as reported by~\citet{SSSGetal_2011}. 

To have a fair assessment, the same procedure for clustering proteins  
is also applied to the proteins identified using protocols of~\citet{SSSGetal_2011}.
The observed protein $P$-value is defined similarly. 
 Database proteins that contain 
  any of the best match peptides, one from each spectrum, 
 form the effective protein set.  
 The observed protein $P$-value is simply the  
  fraction of proteins in the effective protein set 
 that have reported protein $P$-values less than the specified threshold. 
The reported protein $P$-value for the head 
protein of each cluster is obtained by applying 
the iterative procedure (involving uses of Stouffer's formula) described in
 ~\citep{SSSGetal_2011}. In panel B of Figure~\ref{Fig:2}, the reported 
protein $P$-values are plotted against the observed protein $P$-values. 
 The agreement between the reported protein $P$-values and the observed protein $P$-values is not as great as in the peptide case. 
 The protein $E$-value is then obtained by multiplying the protein $P$-value
  by the total number of proteins in the effective protein set. 
  
To construct a PFD curve, it is necessary to estimate the number of 
false identifications at a given significance threshold. 
The number of false identifications can be estimated either by 
using the reported protein $E$-values or the number of
 identifications within the decoy databases. 
 The latter is currently widely used mainly because
  accurate protein $E$-values (or $P$-values) are generally hard to attain.     
To investigate the agreement between the PFD curves obtained using 
 decoy databases and using reasonably accurate protein $P$-values, we use spectra 
 acquired from dataset SN 26 and construct the PFD curves obtained using both approaches.  
The good agreement between our $E$-value based PFD~\citep{Soric_1989} 
and the target-decoy based PFD, 
displayed in panel C of Figure~\ref{Fig:2}, is expected because, 
as shown in panel D of Figure~\ref{Fig:1},  
we have already found that the reported PFD 
and the observed PFD (computed by using a known protein mixture)
 are nearly identical.  
The disagreement between the 
 $E$-value based PFD and the target-decoy based PFD using protocols
 of~\citep{SSSGetal_2011} seems to indicate that the moderate uncertainty 
 in DPV can influence the accuracy of the 
 overall PFD estimate in a substantial manner.  
 
For RAId\_DbS, the agreement between our $E$-value based PFD 
and the target-decoy based PFD is further tested using more 
spectral datasets (SN16-SN25),
 see supplementary Figure S6.  In addition to RAId score, RAId\_aPS allows 
 other scoring functions: XCorr, Hyperscore, and Kscore. 
 For completeness, we plot their corresponding 
    protein $P$-value accuracy assessments in supplementary Figures S7-S9; we also 
    present the agreement tests between their $E$-value based PFDs and the 
    target-decoy based PFDs in supplementary Figures S10-S12.   
 
\subsection{Comparison with other methods}
\label{sec:comp-other}

\begin{figure*}[htbp]
\begin{center}
\vspace*{-0.1in}
\includegraphics[width=0.9\textwidth,angle=0]{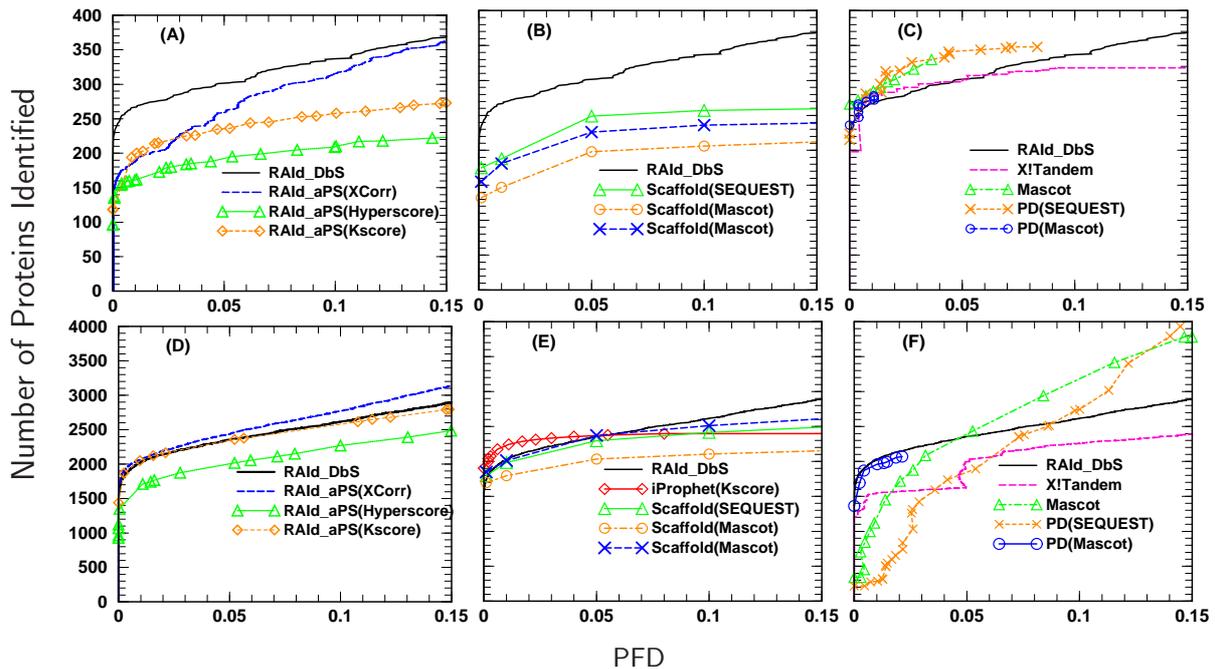}
\vspace*{-0.2in}
\end{center}
\caption{ {\bf Retrieval results of various methods based on their stated PFD values}. Because we can only ensure the accuracy of type-I error 
 of the proposed method,
  this figure only illustrates how our retrieval fares at the stated
 value when compared to other methods. See the text of 
 section~\ref{sec:comp-other} for more details.  
 To avoid clutter, results from using samples of {\it Escherichia coli} whole cell lysyte, SN 27-30, are displayed in three panels (A, B, and C). 
 Within each panel, the results from RAId\_DbS are always shown as a reference curve. Similarly, results from using samples of {\it Homo sapiens} heart cells, SN 31-63, are also displayed in three panels (D, E, and F). 
 The iProphet~\citep{SDLEetal_2011} results in panel E were downloaded from PeptideAtlas 
  instead of being computed.  \vspace*{-0.1in}  
} 
\label{Fig:3}
\end{figure*}

The previous two subsections focus on the accuracy of type-I error control. 
Although it is possible to accurately control type-I error for
 some protein ID methods, this seems not the central focus of 
 all protein ID methods. Many protein ID methods prefer to 
 use the decoy database search results to pragmatically provide statistical significances 
 for retrieval results from the target (organismal) database. When this approach is used, the 
 retrieval results are displayed in terms of a parametric PFD plot: the parameter
  is some kind of significance score used to prioritize the identifications, the 
 abscissa shows the proportion of false discovery and the ordinate displays 
 the number of identifications found in the target database.
 In general, a large number of target identifications at a small PFD value 
  indicates a good retrieval, provided that the number of decoy identifications 
  accurately reflects the number of false identifications in the target database.
   However, one should note that 
   the fulfillment of the aforementioned condition requires accurate type-I error control. 
 Investigating and improving the statistical accuracy of type-I error control 
 of existing protein ID methods is beyond the scope of this paper 
 and we believe that it is best done by developers of individual protein ID 
 software.  
  
To examine how our method compares with others under the pragmatic target-decoy approach, 
we analyze two large datasets from {\it E. Coli} (SN 27-30) and Homo Sapiens (SN 31-63) using
 a variety of protein ID software along with a number of scoring functions. The list of 
 software is given below (with both software version 
 and scoring functions, if given, shown inside a pair of parentheses):  
RAId\_DbS (v. Jan.12.2014; RAId), RAId\_aPS (v. Jan.12.2014; XCorr, Kscore, Hyperscore), 
Mascot (v. 2.4.0, 
{\tt http://www.matrixscience.com/help.html}), 
and X!Tandem (v. 2013.06.15; Hyperscore). 
 The peptide identification software SEQUEST~\citep{EMY_1994} (v. 28) is only used in conjunction with
 other post-processing protein ID software. 
 We list below the post-processing software used  
   (with software version, peptide ID software, 
and peptide scoring functions, if given, shown inside a pair of parentheses): 
iProphet (v. TPP 4.5; X!Tandem; Kscore),
Proteome Discoverer (v. 1.3, 
{\tt http://www.thermofisher.com}\\{\tt /en/home.html};
SEQUEST, Mascot),
 and Scaffold Q+/Q+S (v. 4.0, 
 {\tt http://www.proteomesoftware.com}; SEQUEST, Mascot). 
 The results are displayed in different panels of Figure~\ref{Fig:3}. 
 Before delving into the details of the results, we first provide the
  information relevant to the generation of the results.  

In terms of peptide identification, RAId\_DbS, RAId\_aPS, Mascot, SEQUEST, and X!Tandem 
 used the same parameters: for
{\it Escherichia coli} whole cell lysate, SN 27-30, the precursor ion mass error tolerance 
is $\pm$ 0.033 Da., the product ion mass error tolerance is $\pm$ 0.033 Da.,  
and up to 5 missed cleavages are allowed; for 
 {\it Homo sapiens} heart cells, SN 31-63, the precursor ion mass error tolerance is $\pm$ 1.4 Da.,
  the product ion mass error tolerance is $\pm$ 0.4 Da., and up to 2 missed cleavages are allowed. 

Both X!Tandem and Mascot have built-in protein ID capability, and the target-decoy approach 
was directly applied to estimate the protein level PFD. The peptide ID outputs from SEQUEST
 and Mascot were also further analyzed using Proteome Discoverer
for protein identification and the target-decoy approach was applied to 
 estimate PFD. For iProphet, we did not compute the PFD but downloaded the results for 
 data group 4 from PeptideAtlas. Peptide identification in this case 
 was done using X!Tandem (v. 2009.10.01; Kscore). 

Whenever the decoy peptide search results are available, 
Scaffold computes the PFDs using the target-decoy approach; 
 otherwise, it computes the PFDs using a probabilistic method. 
 In Figure~\ref{Fig:3} three Scaffold 
 PFD curves are displayed, two of which (shown in triangles and circles)
  are from target-decoy approaches. 
%
%
%
%
The protein PFDs under Scaffold were computed by fixing the peptide threshold at 20\% PFD 
with a minimum of 1 evidence peptide per protein.  We observed that changing the peptide 
threshold to lower values had a small effect on the number of proteins identified.  
We thus used the minimum of 1 peptide per protein to maintain consistency across all methods. 
For RAId\_DbS and RAId\_aPS, the PFD estimates do not require 
user-added target-decoy methods.  
RAId\_DbS and RAId\_aPS compute the PFDs using the Sori\'c formula~\citep{Soric_1989}.

 Examinations of different panels of Figure~\ref{Fig:3} indicates that 
 the retrieval efficacy of the proposed method (shown in RAId\_DbS and RAId\_aPS 
  PFD curves) is comparable with existing protein ID methods, 
 even though only at the stated values. However, it should be noted that 
  the proposed method does have a few advantages. 
  First, it reports accurate protein $P$-values, providing 
  accurate type-I error control. Second, the PFD curves obtained using this method
   show stability across different mass resolution requirement and data sets, 
   while some methods seem to exhibit fluctuations of notable amplitudes. 
 \vspace*{-0.2in}

\section{Discussion}
Our investigation indicates that it is possible to achieve 
faithful protein $P$-value assignment, hence accurate type-I error control, in
 protein identifications. Since our approach is founded on a 
 derived mathematical formula that requires accurate peptide 
 $E$-values as input, it is evident that accurate protein $P$-values 
 require accurate statistical significance at the peptide
 identification level. 

 The discrepancy between the computed protein $P$-value and the PFD results
  in our implementation of the method of~\citet{SSSGetal_2011} is interesting.
 Based on the results in Fig.~\ref{Fig:2}, the peptide $P$-values are reasonably 
  accurate albeit exhibiting slightly larger fluctuations than the 
  results from RAId\_DbS.    
 In addition to the possibility of accumulating uncertainty of 
  peptides' $P$-values, the other possibility is that the iterative procedure
   to choose the combination yielding the most significant Z-score
 may skew the $P$-values towards the significant side.
 Investigation of the origin of the PFD and $P$-value discrepancy when using the method of~\citet{SSSGetal_2011}, however, is beyond the scope of the current study and 
 might be most appropriately done by the authors of~\citep{SSSGetal_2011}.

As explained earlier, we allow more than one candidate peptide per spectrum to
 accommodate the possibility of peptide co-elution. However, 
 readers may ask why do we choose to use DPVs for lower-ranking
 peptides per spectrum instead of using ordered statistics. The reason is that 
  in this context using ordered statistics beyond the first is not meaningful:
 the $n$th ordered statistics assumes that for a given query spectrum 
 the best $n-1$ scored  peptides are spurious while the rank-$n$ peptide is 
  the underlying peptide whose fragmentation yields the query spectrum. 
  This contradicts the general idea of using a scoring function: 
   among candidate peptides of a query spectrum, the better a peptide 
   scores the more likely it is the underlying peptide. 
  On the other hand, when using the DPV for the rank-$n$ peptide, 
  we are essentially assuming that the top $n-1$ candidate peptides of the query 
  spectrum are co-eluted underlying peptides and are not considered to 
  be spurious.

The protein identification method proposed in this paper 
illustrates the possibility of accurate type-I error control, 
 providing a theoretically sound significance assignment method 
 that is also pragmatically simpler than the target-decoy approach.  
This is particularly important since  
 the number of identified proteins versus PFDs provides trustworthy 
  retrieval results only if the reported PFDs
  truly reflects the proportion of false discoveries. 
 Evidently, to achieve accurate type-I error control is a task 
 best done by developers of individual software. Only when this is 
  accomplished can a true retrieval comparison among different methods
   be done.

 Since we did not focus on type-II error, there is definite room for improvement
  in terms of retrieval efficacy. We note that the information of negatives 
  (segments of a candidate protein not covered by the protein's 
  evidence peptides) is not  used. 
  We also believe that, in principle, scoring functions for peptide identification 
   can also be improved to better separate true underlying peptides from false positives. 
 Currently, we are using a flat peptide weight (by the number of proteins a peptide
  covers). It is perceivable that more sophisticated weighting may be useful in 
   better separating true positive proteins from false positives.       
 It is our plan to investigate these avenues of improvement 
 in the near future.   
 \vspace*{-0.24in}

%
%


\section*{Acknowledgement}

\paragraph{Funding\textcolon} 
This work was supported by the Intramural Research Program of the National Library of
Medicine at the National Institutes of Health.


\clearpage
\setcounter{figure}{0}
\renewcommand{\thefigure}{S\arabic{figure}}
\renewcommand{\thetable}{S\arabic{table}}

\onecolumn
\section*{Supplementary Information}

\subsection*{Example application of formula (3) of the main text}
To illustrate the use of the formula used to compute a protein $P$-value, 
let us consider the following toy example.
Let protein $\Pi$ have six peptide evidences $\pi_1,~\pi_2,\ldots,\pi_6$ 
that falls into three groups $\{ \pi_1 \}$, $\{\pi_2, \pi_3\}$, and $\{ \pi_4,\pi_5,\pi_6\}$, respectively with weights $1$, $1/3$, $1/2$. This means that 
 peptide $\pi_1$ is a subsequence of protein $\Pi$ only, peptides $\pi_2$ and $\pi_3$ 
  are subsequences of three proteins ($\Pi$ and two others), and peptides 
  $\pi_4$, $\pi_5$, and $\pi_6$ are subsequences of two proteins ($\Pi$ and another one). Let the peptide $E$-values of these six evidence peptides be $e_1,~e_2,\ldots,~e_6$, all less than one. Under our approximating database $P$-value by $E$-value, this means that the evidence peptides have their respective database $P$-values $e_1,~e_2,\ldots,~e_6$.  From the information above we know that $m=3$, $n_1=1$, $n_2=2$, $n_3=3$, $w_1=1$, $w_2=1/3$, $w_3=1/2$, $r_1 = 1/w_1 =1$, 
  $r_2=1/w_2=3$, and $r_3=1/w_3=2$.  

To use formula (3) in the manuscript to compute a protein $P$-value, 
 we first need the quantity $\tau$ given by eq.~(1) in the main text. 
 In the current case-- $m=3$, $n_1=1$, $n_2=2$, $n_3=3$ and with peptides database $P$-values $e_1,~e_2,\ldots,~e_6$ -- the quantity $\tau$ can be written as 
\begin{equation}
\tau = \prod_{k=1}^{3} \left[ \prod_{j=1}^{n_k} e_{k;j} \right]^{w_k} = e_1^{w_1} e_2^{w_2}e_3^{w_2}e_4^{w_3}e_5^{w_3}e_6^{w_3}. \nonumber
\end{equation}      
And the protein $P$-value is given by
\begin{equation} 
F(\tau) = \left[ \prod_{l=1}^{3}  r_l^{n_l} \right] \sum_{k=1}^3 \,
\sum_{\mathcal{G}(k)}  
\;\biggr\{
\frac{1}{r_k^{g_k+1}}  H(- r_k \ln \tau \, ,\, g_k) \times  
 \biggr(  \prod_{j=1,j\ne k}^{3}  \frac{ (n_j-1+g_j)!}{(n_j-1)!  g_j!}  \frac{(-1)^{g_j}}{(r_j-r_k)^{n_j+g_j}}  \biggr) \biggr\}  
\label{eq:1}
,
\end{equation}
where $H(- r_k \ln \tau \, ,\, g_k)$ is given by
\begin{equation}
H(-r_k \ln \tau \, ,\, g_k) =  \exp( r_k \ln \tau) \sum_{k=0}^{g_k} \frac{ (-r_k \ln \tau)^{k}}{k!}. 
\nonumber
\end{equation}
Remember that $\sum_{\mathcal{G}(k)}$ enumerates each set of
 nonnegative integers $\{g_1,g_2,\ldots,g_m\}$ that satisfies  
 the $k$-dependent constraint $\sum_{i=1}^m g_i = n_k-1$, it is thus 
 possible to replace the $\sum_{\mathcal{G}(k)}$ in eq.~({\ref{eq:1}) by 
an $m$-dimensional summation with an explicit constraint. 
Specifically, we can rewrite the sum over set as 
\begin{eqnarray} 
F(\tau) &=& \left[ \prod_{l=1}^{3}  r_l^{n_l} \right] \sum_{k=1}^3 \,
\sum_{g_1=0}^{n_k-1}  \sum_{g_2=0}^{n_k-1}   \sum_{g_3=0}^{n_k-1} 
\delta_{g_1+g_2+g_3,n_k-1}   
\;\biggr\{
\frac{1}{r_k^{g_k+1}}  H(- r_k \ln \tau \, ,\, g_k) \times   \nonumber \\
&& \times \biggr(  \prod_{j=1,j\ne k}^{3}  \frac{ (n_j-1+g_j)!}{(n_j-1)!  g_j!}  \frac{(-1)^{g_j}}{(r_j-r_k)^{n_j+g_j}}  \biggr) \biggr\}. 
\label{eq:2}
\end{eqnarray}
where the Kronecker delta function $\delta_{n,n'}$ takes value one if 
$n=n'$ but zero otherwise. 

The first product on the right hand side of eq.~(\ref{eq:2}) is equal to
\begin{equation} 
\left[ \prod_{l=1}^{3}  r_l^{n_l} \right] =
 (r_1)^1 \cdot (r_2)^2 \cdot (r_3)^3 
= 1^1 \cdot 3^2 \cdot 2^3 \,.
\nonumber
\label{eq:3}
\end{equation}
After this overall factor is obtained, the main task is to evaluate the summation over $k$, which ranges from $k=1$ to $k=3$. For each $k$, we are only interested in the non-negative integral $g_i$s that satisfy the $k$-dependent constraint 
$\sum_{i=1}^{3} g_i = n_k - 1$.  When $k=1$ we have $n_1 = 1$ 
 and the constrained summation 
\begin{equation} 
 \sum_{g_1=0}^{0}  \sum_{g_2=0}^{0}   \sum_{g_3=0}^{0} \delta_{g_1+g_2+g_3,0}
\nonumber 
\end{equation}
only allows one valid \{$g_1,g_2,g_3$\} set, namely, \{0,0,0\}. 
For $k=2$ we have $n_2 = 2$ and the constrained summation 
\begin{equation} 
 \sum_{g_1=0}^{1}  \sum_{g_2=0}^{1}   \sum_{g_3=0}^{1} \delta_{g_1+g_2+g_3,1}
\nonumber 
\end{equation}
allows three sets of valid \{$g_1,g_2,g_3$\}, namely, \{1,0,0\}, \{0,1,0\} and \{0,0,1\}.  For $k=3$ we have $n_3 = 3$ and the constrained summation  
\begin{equation} 
 \sum_{g_1=0}^{2}  \sum_{g_2=0}^{2}   \sum_{g_3=0}^{2} \delta_{g_1+g_2+g_3,2}
\nonumber 
\end{equation}
allows six sets of valid \{$g_1,g_2,g_3$\}, namely, 
\{1,1,0\}, \{1,0,1\}, \{0,1,1\}, \{2,0,0\}, \{0,2,0\} and  \{0,0,2\}.
Each valid set of $\{g_1,g_2,g_3\}$ must be substituted into the summand 
(inside the pair of curly braces) of 
eq.~(\ref{eq:2}) to yield its respective contribution for the $P$-value.

\clearpage
\subsection*{Supplementary Tables}
\begin{table*}[!hb]
\begin{center}
\caption{ {\bf MS/MS data group 1}.  The corresponding spectral datasets
 are produced by using Sigma49 (a protein standard mixture composed of 49 known human proteins), and are downloaded from the National Center for Biotechnology Information (Peptidome database) at \href{ftp://ftp.ncbi.nih.gov/pub/peptidome/studies/PSEnnn/PSE108}{ftp://ftp.ncbi.nih.gov/pub/peptidome/studies/PSEnnn/PSE108}. 
The column heading SN represents sample index, the abbreviation CGL stands for
  chromatography gradient length, and the column heading $n_s$ stands for 
  the number of spectra acquired. The rest of the column headings 
   are self-explanatory. 
\vspace{0.2cm}}
\label{UPS1:DATASET}
\begin{threeparttable} 
\begin{tabular}{|c|c|c|c|c|l| } \hline
SN &  Sample Load  & Instrument & CGL(minutes) & $n_s$ & File Name \\ \hline
1  & 5 fmol    &	LTQ Orbitrap  &	45   &  1,531 & PSM1027\_07FEB15\_ABRF\_FT\_5a.mzXML \\ \hline	
2  & 5 fmol    &	LTQ Orbitrap  &	45	 &  1,902 & PSM1028\_07FEB15\_ABRF\_FT\_5b.mzXML  \\ \hline	
3  & 5 fmol    &	LTQ Orbitrap  &	45	 &  2,014 & PSM1029\_07FEB15\_ABRF\_FT\_5c.mzXML \\ \hline	
4  & 10 fmol   &	LTQ Orbitrap  &	45	 &  2,026 & PSM1027\_07FEB15\_ABRF\_FT\_10a.mzXM\\ \hline	
5  & 10 fmol   &	LTQ Orbitrap  &	45	 &  2,125 & PSM1028\_07FEB15\_ABRF\_FT\_10b.mzXML\\ \hline	
6  & 10 fmol   &	LTQ Orbitrap  &	45	 &  2,253 & PSM1029\_07FEB15\_ABRF\_FT\_10c.mzXML \\ \hline	
7  & 25 fmol   &	LTQ Orbitrap  &	45	 &  2,772 & PSM1027\_07FEB15\_ABRF\_FT\_25a.mzXML \\ \hline	
8  & 25 fmol   &	LTQ Orbitrap  &	45	 &  2,669 & PSM1028\_07FEB15\_ABRF\_FT\_25b.mzXML \\ \hline	
9  & 25 fmol   &	LTQ Orbitrap  &	45	 &  2,504 & PSM1029\_07FEB15\_ABRF\_FT\_25c.mzXML \\ \hline	
10 & 50 fmol   &	LTQ Orbitrap  &	45	 &  3,259 & PSM1027\_07FEB15\_ABRF\_FT\_50a.mzXML \\ \hline	
11 & 50 fmol   &	LTQ Orbitrap  &	45	 &  3,406 & PSM1028\_07FEB15\_ABRF\_FT\_50b.mzXML \\ \hline	
12 & 50 fmol   &	LTQ Orbitrap  &	45	 &  2,993 & PSM1029\_07FEB15\_ABRF\_FT\_50c.mzXML \\ \hline	
13 & 100 fmol   & LTQ Orbitrap	  &	 45	 &  3,629 & PSM1027\_07FEB15\_ABRF\_FT\_100a.mzXML \\ \hline	
14 & 100 fmol   & LTQ Orbitrap	  &	 45	 &  3,622 & PSM1028\_07FEB15\_ABRF\_FT\_100b.mzXML \\ \hline	
15 & 100 fmol   & LTQ Orbitrap  & 45	 &  3,592 & PSM1029\_07FEB15\_ABRF\_FT\_100c.mzXML \\ \hline	
\end{tabular} 
\end{threeparttable}
\end{center}
\end{table*}  

\begin{table*}[!hb]
\begin{center}
\caption{ {\bf MS/MS data group 2}.  The corresponding spectral datasets are produced by using a complex protein mixture of {\it Escherichia coli K-12}  whole cell 
lysate, and 
are downloaded from the Pacific Northwest National Laboratory at \href{http://omics.pnl.gov/}{http://omics.pnl.gov/}.
\vspace{0.2cm}
\label{PNNL:DATASET}}
\begin{threeparttable} 
\begin{tabular}{|c|c|c|c|l| } \hline
SN & Instrument & CGL(minutes) & $n_s$ & File Name \\ \hline
16 & 	LTQ  Orbitrap    & 100 & 18,573 & Ecoli432\_R1-rr\_18Dec09\_Falcon\_09-09-14.mzXML \\ \hline
17 & 	LTQ  Orbitrap    & 100 & 18,585 & Ecoli432\_R2\_7Dec09\_Falcon\_09-09-15.mzXML \\ \hline
18 & 	LTQ  Orbitrap    & 100 & 18,669 & Ecoli432\_R3\_7Dec09\_Falcon\_09-09-16.mzXML \\ \hline
19 & 	LTQ  Orbitrap    & 100 & 18,585 & Ecoli432\_R4\_15Dec09\_Falcon\_09-09-16.mzXML \\ \hline
20 & 	LTQ  Orbitrap    & 100 & 18,650 & Ecoli433\_R1\_7Dec09\_Falcon\_09-09-14.mzXML \\ \hline
21 & 	LTQ  Orbitrap 	 & 100 & 18,763 & Ecoli433\_R2\_7Dec09\_Falcon\_09-09-15.mzXML \\ \hline
22 & 	LTQ  Orbitrap 	 & 100 & 18,770 & Ecoli433\_R4\_13Dec09\_Falcon\_09-09-16.mzXML \\ \hline
23 & 	LTQ  Orbitrap    & 100 & 18,488 & Ecoli434\_R1\_7Dec09\_Falcon\_09-09-14.mzXML \\ \hline
24 & 	LTQ  Orbitrap 	 & 100 & 18,923 & Ecoli434\_R2\_7Dec09\_Falcon\_09-09-15.mzXML \\ \hline
25 & 	LTQ  Orbitrap 	 & 100 & 19,010 &  Ecoli434\_R3\_7Dec09\_Falcon\_09-09-16.mzXML\\ \hline
26 & 	LTQ  Orbitrap 	 & 100 & 18,737 & Ecoli434\_R4\_13Dec09\_Falcon\_09-09-16.mzXML \\ \hline
\end{tabular} 
\end{threeparttable}
\end{center}
\end{table*}

\begin{table*}[!hb]
\begin{center}
\caption{ {\bf MS/MS data group 3}. The corresponding spectral datasets are produced by using a complex protein mixture of {\it Escherichia coli}  whole cell 
lysate prepared in house.  
\vspace{0.2cm}
\label{INHOUSE:DATASET}}
\begin{threeparttable} 
\begin{tabular}{|c|c|c|c|l| } \hline
SN & Instrument & CGL(minutes) & $n_s$ & File Name \\ \hline
27 & Orbitrap Elite   & 90  &  24,280  &  E\_L\_2.mzML \\ \hline
28 & Orbitrap Elite   & 90 & 22,435 &  E\_M\_2.mzML \\ \hline
29 & Orbitrap Elite   & 90 & 23,875 &  E\_H\_2.mzML \\ \hline
30 & Orbitrap Elite   & 90 & 18,573 &  E\_S\_2.mzML \\ \hline
\end{tabular} 
\end{threeparttable}
\end{center}
\end{table*}

\begin{table*}[!hb]
\begin{center}
\caption{ {\bf MS/MS data group 4}. 
 The corresponding spectral datasets are produced by using 
 SDS-PAGE protein fractionation extraction of human lung cells, and are
  downloaded from PeptideAtlas database at \href{ftp://ftp.peptideatlas.org/pub/PeptideAtlas/Repository/PAe001771}{ftp://ftp.peptideatlas.org/pub/PeptideAtlas/Repository/PAe001771}.
\vspace{-0.2cm}
\label{INHOUSE:DATASET}}
\begin{threeparttable} 
\begin{tabular}{|c|c|c|c|l| } \hline
SN & Instrument & CGL(minutes) & $n_s$ & File Name \\ \hline
31-63 & LTQ Orbitrap & 120  &  340,861   &  Roche\_human\_lung\_001.mzML - Roche\_human\_lung\_033.mzML \\ \hline
\end{tabular} 
\end{threeparttable}
\end{center}
\end{table*}


\clearpage
\subsection*{Supplementary Figures}

\begin{figure*}[htbp]
\begin{center}
\includegraphics[width=0.92\textwidth,angle=0]{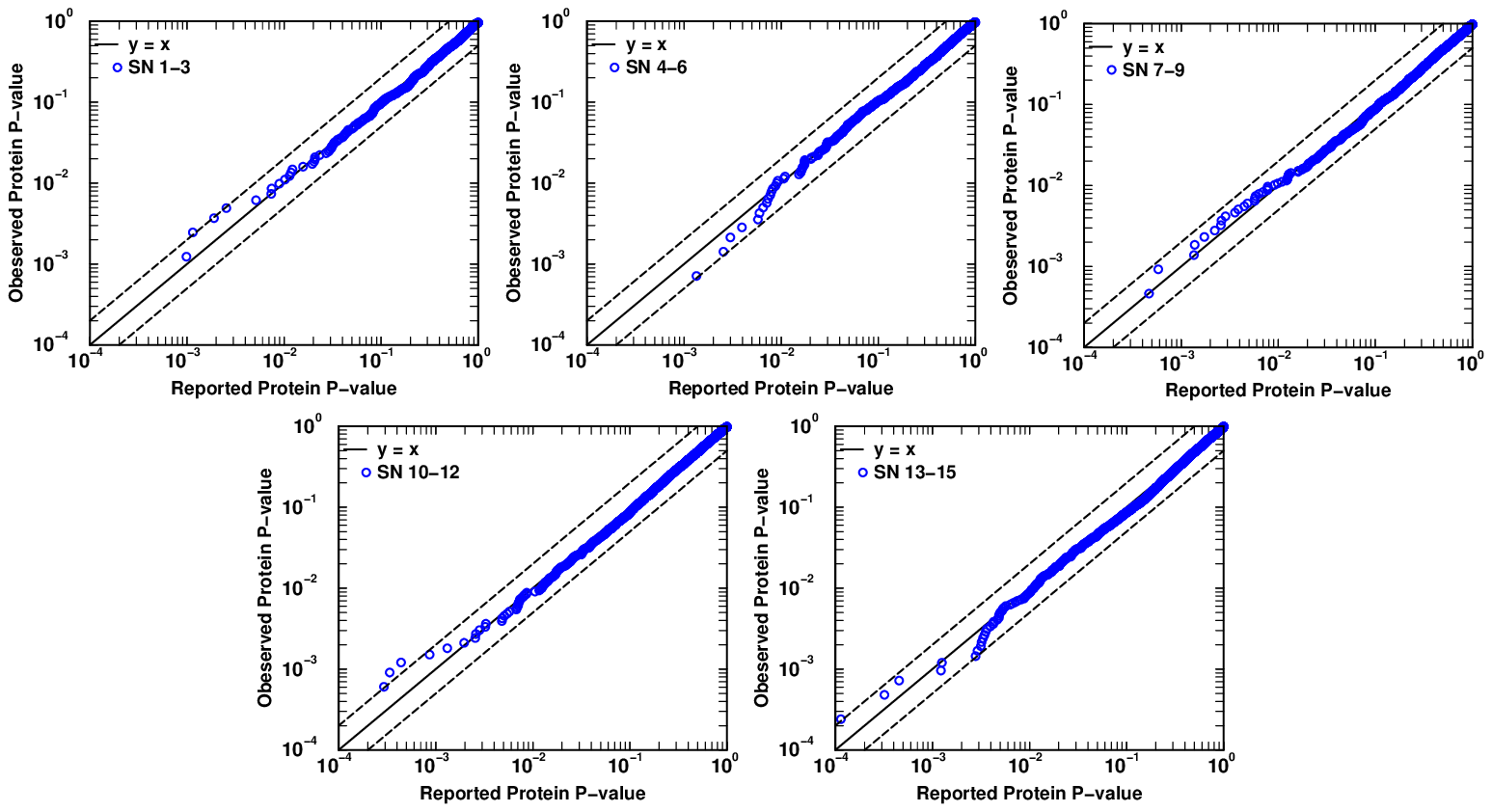}
\end{center}
\caption{ {\bf Accuracy assessment of the protein {\it P}-value  reported
 by the proposed method using data group 1 and RAId\_DbS's peptide $E$-values}. 
 All spectra in data group 1 are used to search the {\it Homo sapiens} database
  with precursor-ion mass tolerance $\pm$ 0.033 Da., product-ion mass tolerance $\pm$ 0.8 Da. and up-to-2 missed cleavage sites allowed. The accuracy of the reported protein {\it P}-value  is evaluated by plotting it versus the observed protein {\it P}-value, see main text for details. The closer the above curves are to the $ y = x $ line the more accurate are the reported  protein {\it P}-values. The two dash lines, $y = 2x$ and $y = 0.5x$, are provided as visual guides regarding how close/off the reported protein {\it P}-values are to the $ y = x $ line.} 
\label{Fig:S1}
\end{figure*} 

\clearpage

\begin{figure*}[htbp]
\begin{center}
\includegraphics[width=0.92\textwidth,angle=0]{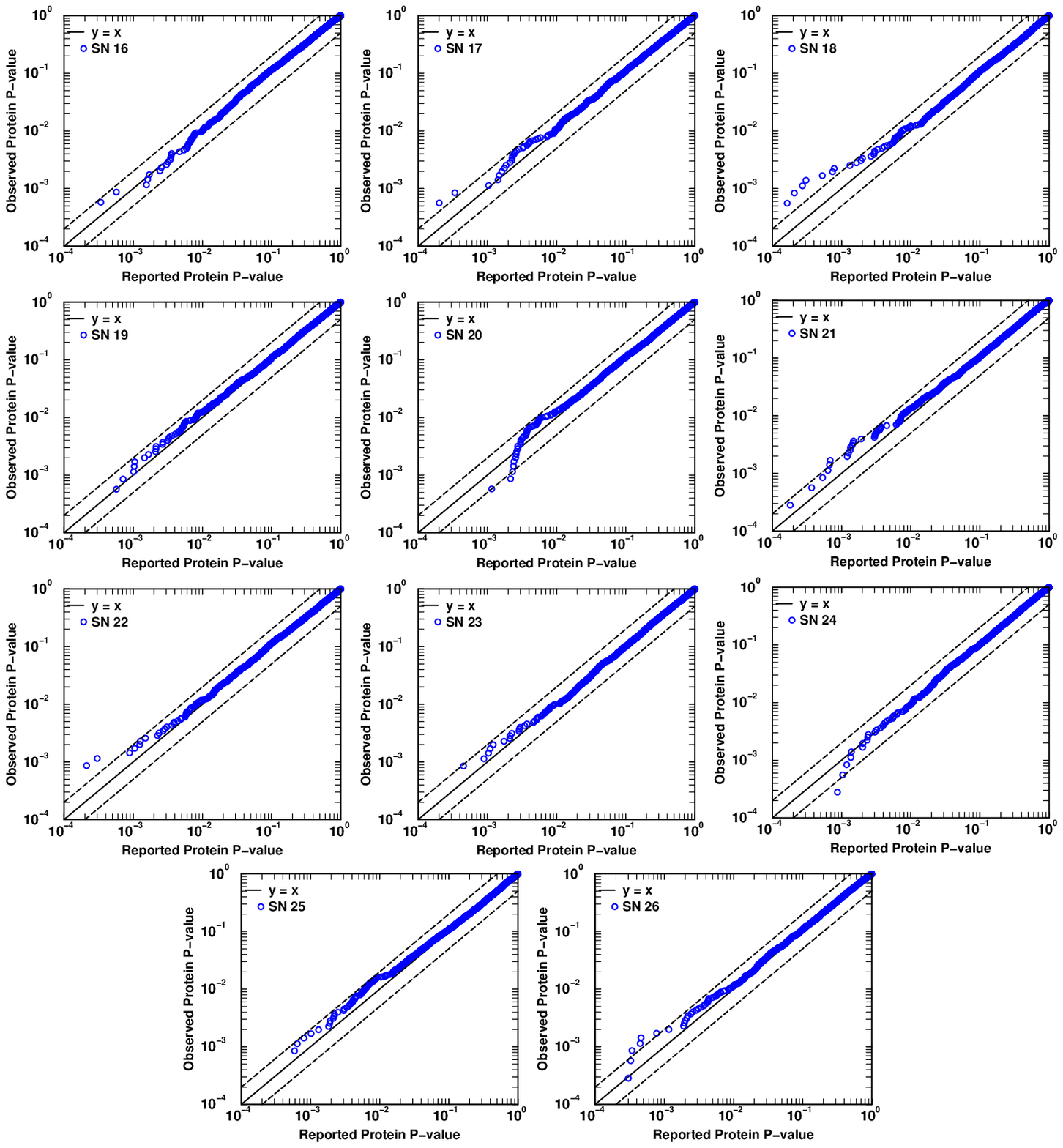}
\end{center}
\caption{ {\bf Accuracy assessment of the protein {\it P}-value  reported
 by the proposed method using data group 2 and RAId\_DbS's peptide $E$-values}. 
 All spectra in data group 2 are used to search the {\it Escherichia coli} database
  with precursor-ion mass tolerance $\pm$ 0.033 Da., product-ion mass tolerance 
  $\pm$ 0.033 Da. and up-to-2 missed cleavage sites allowed. The accuracy of the reported protein {\it P}-value  is evaluated by plotting it versus the observed protein {\it P}-value, see main text for details. The closer the above curves are to the $ y = x $ line the more accurate are the reported  protein {\it P}-values. The two dash lines, $y = 2x$ and $y = 0.5x$, are provided as visual guides regarding how close/off the reported protein {\it P}-values are to the $ y = x $ line.
} 
\label{Fig:S2}
\end{figure*} 

\clearpage

\begin{figure*}[htbp]
\begin{center}
\includegraphics[width=0.92\textwidth,angle=0]{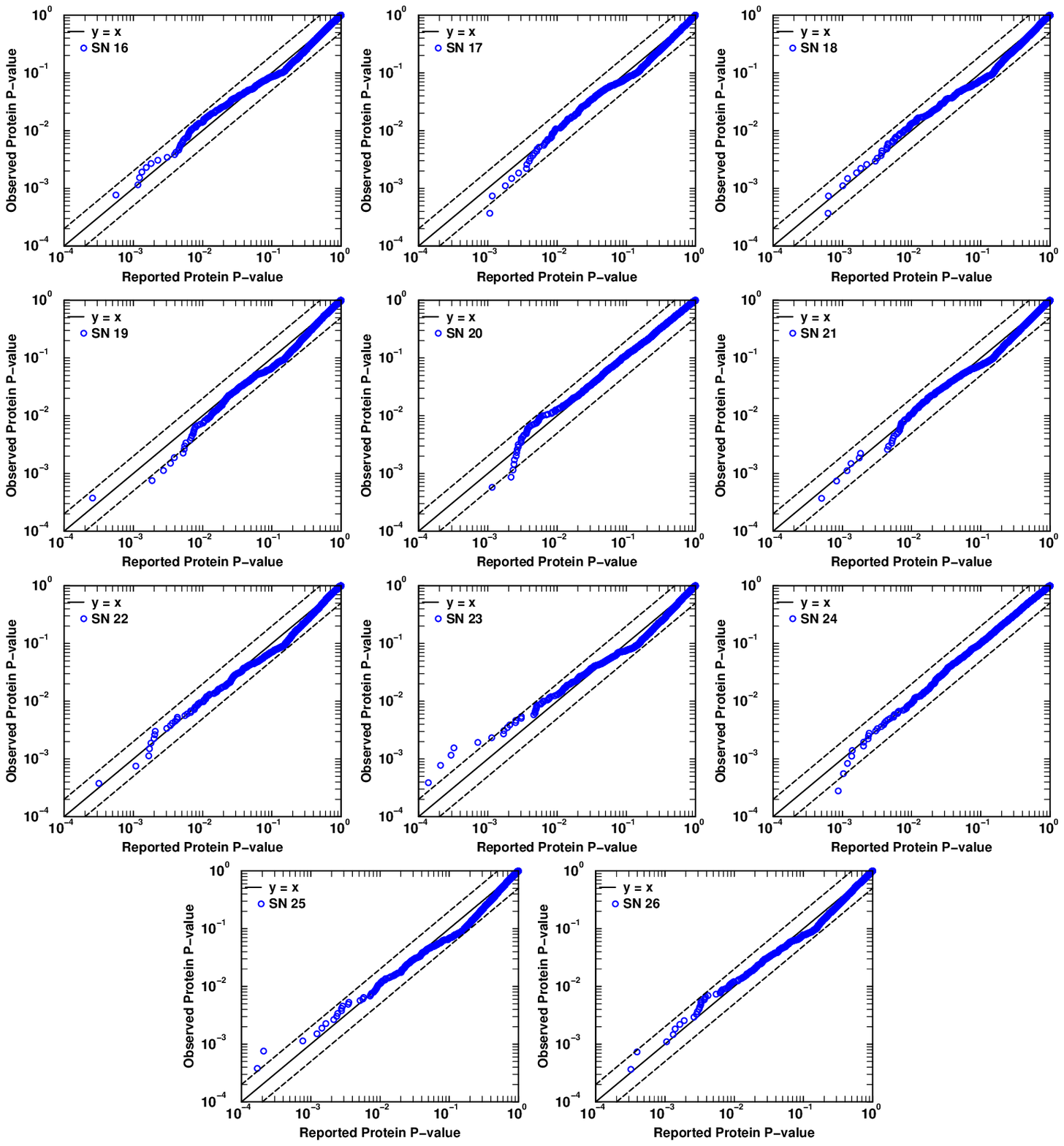}
\end{center}
\caption{ {\bf Accuracy assessment of the reported protein {\it P}-value  
 by the proposed method when the mass tolerances are very small.}
 The protein $P$-values are obtained by using RAId\_DbS's peptide $E$-values.  
 All spectra in data group 2 are used to search the {\it Escherichia coli} database
  with precursor-ion mass tolerance $\pm$ 0.0033 Da., product-ion mass tolerance 
  $\pm$ 0.0033 Da. and up-to-2 missed cleavage sites allowed. The accuracy of the reported protein {\it P}-value  is evaluated by plotting it versus the observed protein {\it P}-value, see main text for details. The closer the above curves are to the $ y = x $ line the more accurate are the reported  protein {\it P}-values. The two dash lines, $y = 2x$ and $y = 0.5x$, are provided as visual guides regarding how close/off the reported protein {\it P}-values are to the $ y = x $ line.
  } 
\label{Fig:S3}
\end{figure*} 

\clearpage

\begin{figure*}[htbp]
\begin{center}
\includegraphics[width=0.92\textwidth,angle=0]{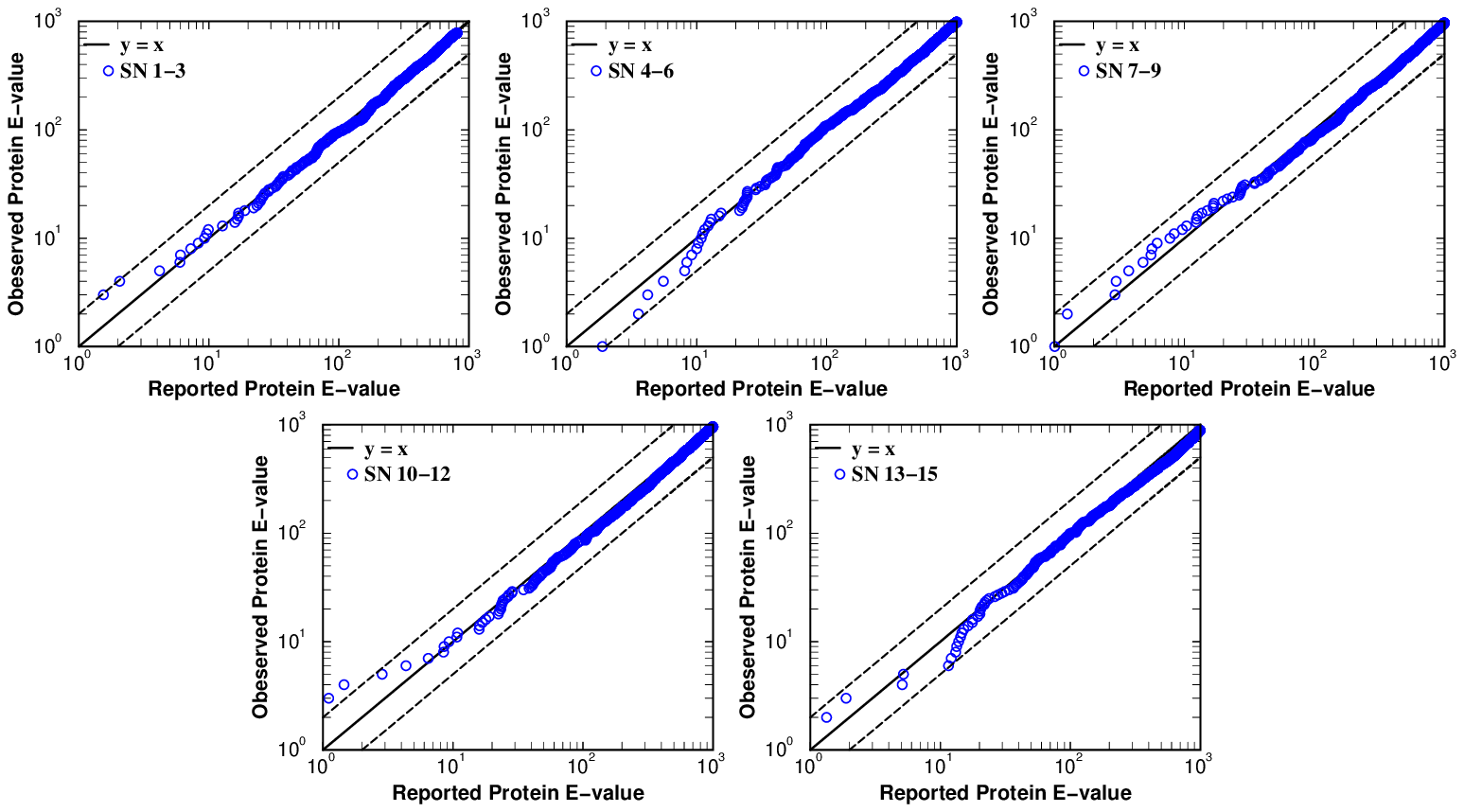}
\end{center}
\caption{{\bf Accuracy assessment of the protein {\it E}-value  reported
 by the proposed method using data group 1 and RAId\_DbS's peptide $E$-values}. 
 All spectra in data group 1 are used to search the {\it Homo sapiens} database
  with precursor-ion mass tolerance $\pm$ 0.033 Da., product-ion mass tolerance $\pm$ 0.8 Da. and up-to-2 missed cleavage sites allowed. The accuracy of the reported protein {\it P}-value  is evaluated by plotting it versus the observed protein {\it P}-value, see main text for details. The closer the above curves are to the $ y = x $ line the more accurate are the reported  protein {\it E}-values. The two dash lines, $y = 2x$ and $y = 0.5x$, are provided as visual guides regarding how close/off the reported protein {\it E}-values are to the $ y = x $ line.
  } 
\label{Fig:S4}
\end{figure*} 

\clearpage

\begin{figure*}[htbp]
\begin{center}
\includegraphics[width=0.92\textwidth,angle=0]{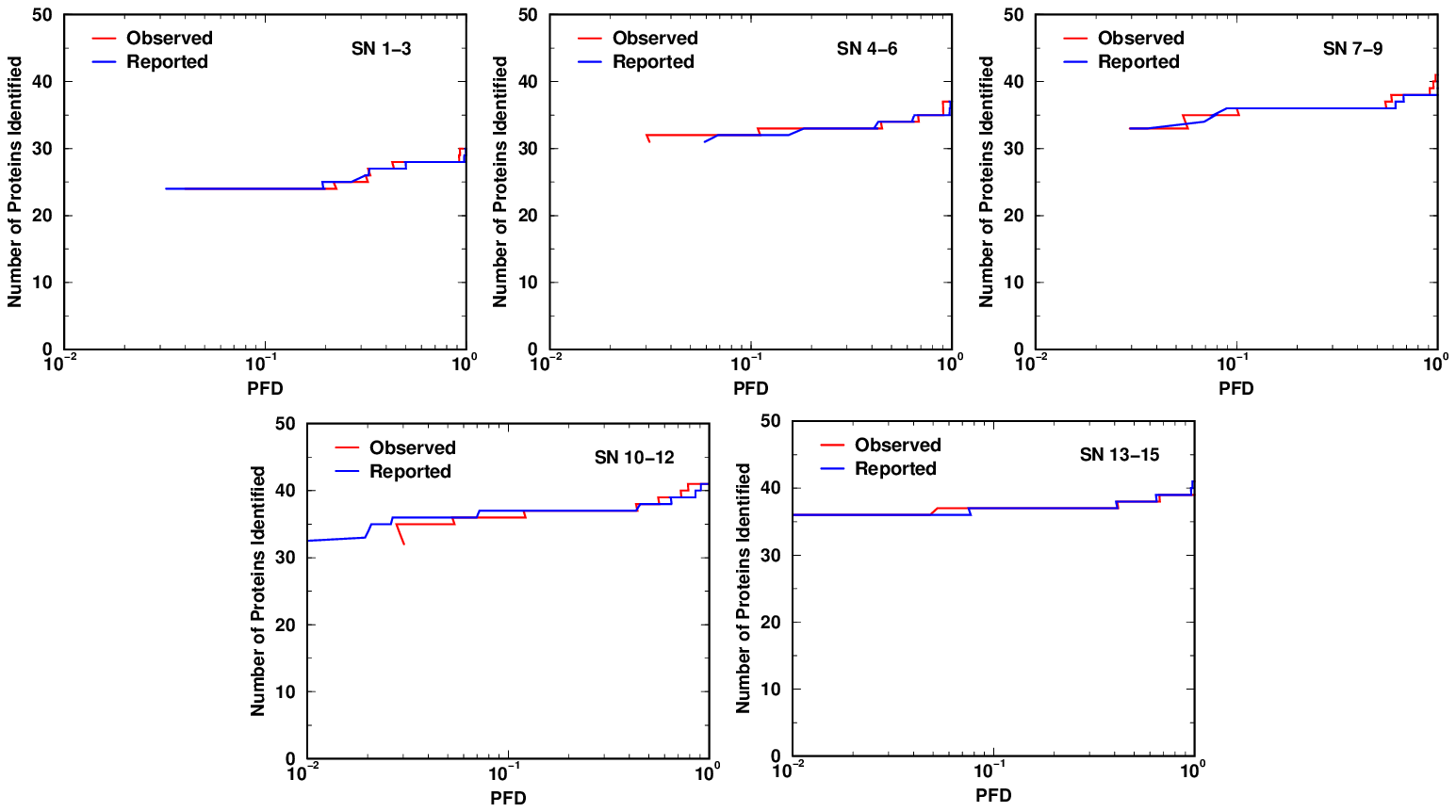}
\end{center}
\caption{ {\bf Accuracy assessment of reported protein PFD 
using data group 1 and RAId\_DbS's peptide $E$-values}. 
 All spectra in data group 1 (MS/MS spectra from SN 1-15) 
 are used to search the {\it Homo sapiens} database
  with precursor-ion mass tolerance $\pm$ 0.033 Da., product-ion mass tolerance $\pm$ 0.8 Da. and up-to-2 missed cleavage sites allowed. The reported PFD is computed by using the reported protein {\it E}-value in Sori\'c's formula, while the observed PFD is obtained from the ratio of the number of false discoveries (false positives) to the total number of discoveries (true positives + false positives) at 
  a given {\it E}-value cutoff.  The closer the observed and reported curves are to each other the better.} 
\label{Fig:S5}
\end{figure*} 

\clearpage

\begin{figure*}[htbp]
\begin{center}
\includegraphics[width=0.95\textwidth,angle=0]{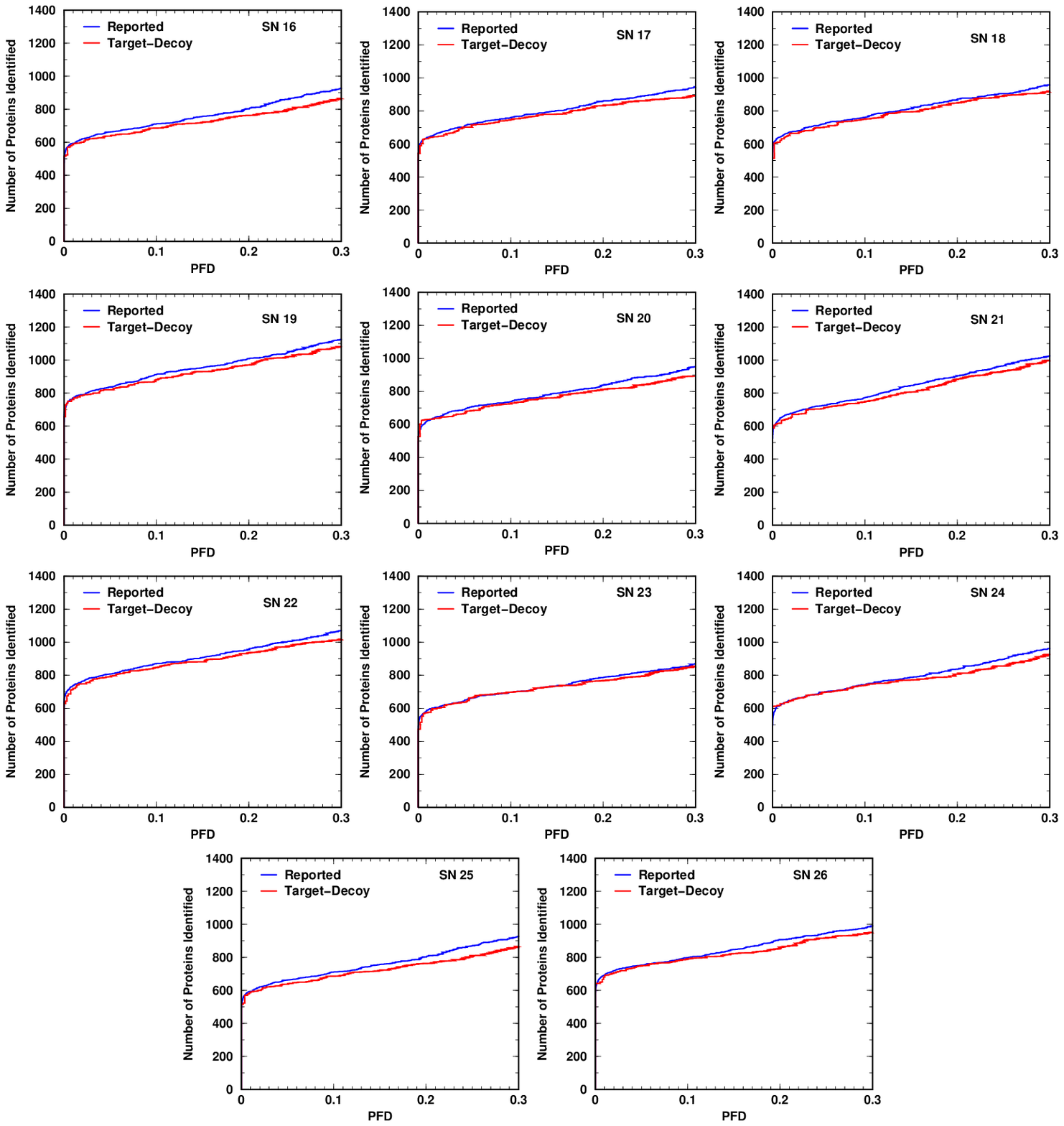}
\end{center}
\caption{  {\bf The agreement between the $E$-value based PFD and 
 the target-decoy based PFD when the peptide $E$-values are from 
 RAId\_DbS}. 
 All spectra in data group 2 (MS/MS spectra SN16-26) 
 are used to search the {\it Escherichia coli} database
  with precursor-ion mass tolerance $\pm$ 0.033 Da., product-ion mass tolerance 
  $\pm$ 0.033 Da. and up-to-2 missed cleavage sites allowed. The $E$-value based 
  PFD is computed by using the reported protein {\it E}-value in Sori\'c's formula, while the target-decoy based PFD is obtained from the ratio of the number of 
   identified decoy proteins to the total number of identified proteins (target + decoy) for a given {\it E}-value cutoff.  Within each panel, the closer 
  the two curves are to each other the better.}
\label{Fig:S6}
\end{figure*} 

\clearpage

\begin{figure*}[htbp]
\begin{center}
\includegraphics[width=0.92\textwidth,angle=0]{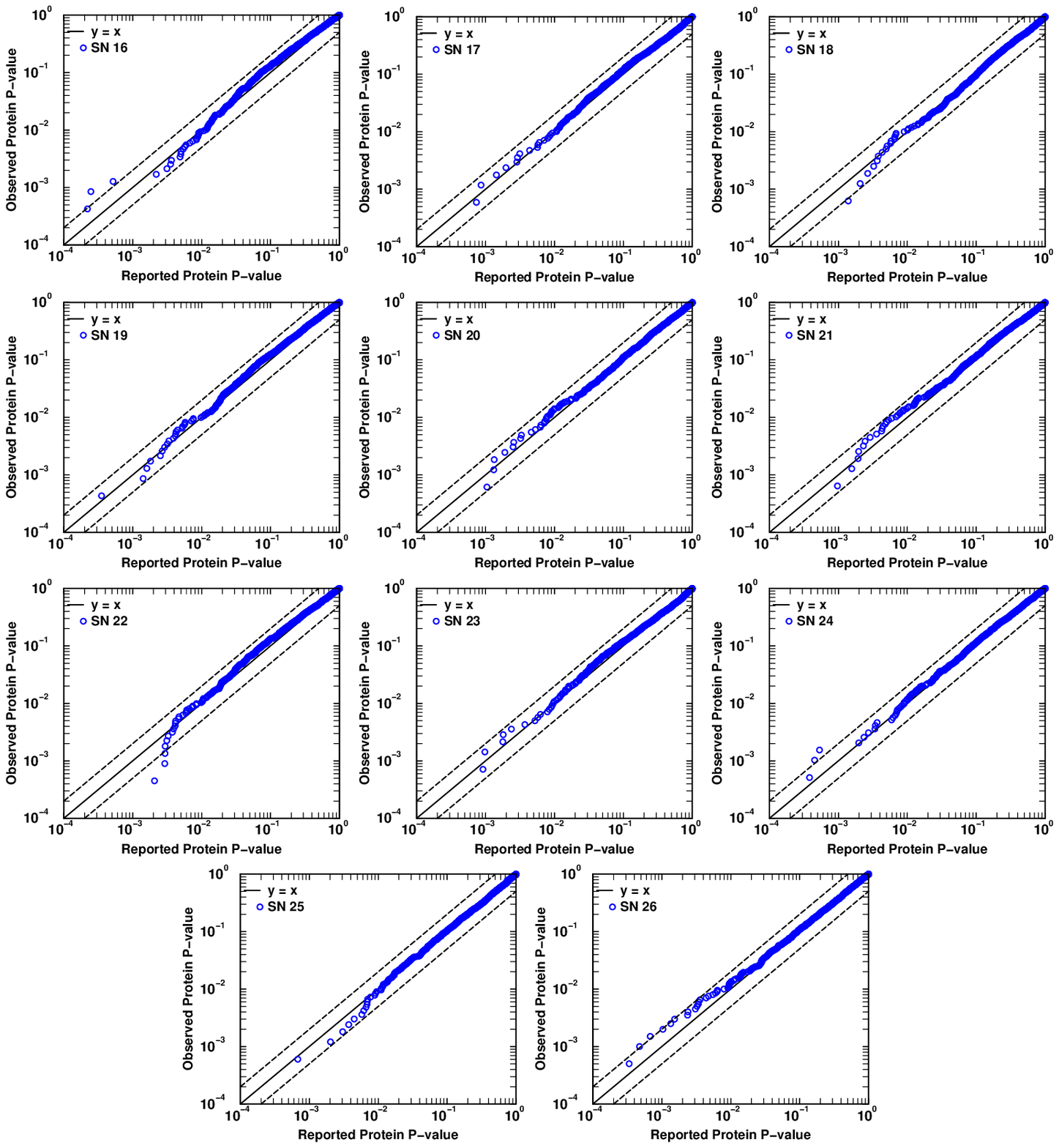}
\end{center}
\caption{ 
{\bf Accuracy assessment of the protein {\it P}-value 
  using data group 2 and RAId\_aPS's peptide $E$-values
  when the selected scoring function is XCorr}. 
 All spectra in data group 2 are used to search the {\it Escherichia coli} database
  with precursor-ion mass tolerance $\pm$ 0.033 Da., product-ion mass tolerance 
  $\pm$ 0.033 Da. and up-to-2 missed cleavage sites allowed. The accuracy of the reported protein {\it P}-value  is evaluated by plotting it versus the observed protein {\it P}-value, see main text for details. The closer the above curves are to the $ y = x $ line the more accurate are the reported  protein {\it P}-values. The two dash lines, $y = 2x$ and $y = 0.5x$, are provided as visual guides regarding how close/off the reported protein {\it P}-values are to the $ y = x $ line.} 
\label{Fig:S7}
\end{figure*} 

\clearpage

\begin{figure*}[htbp]
\begin{center}
\includegraphics[width=0.92\textwidth,angle=0]{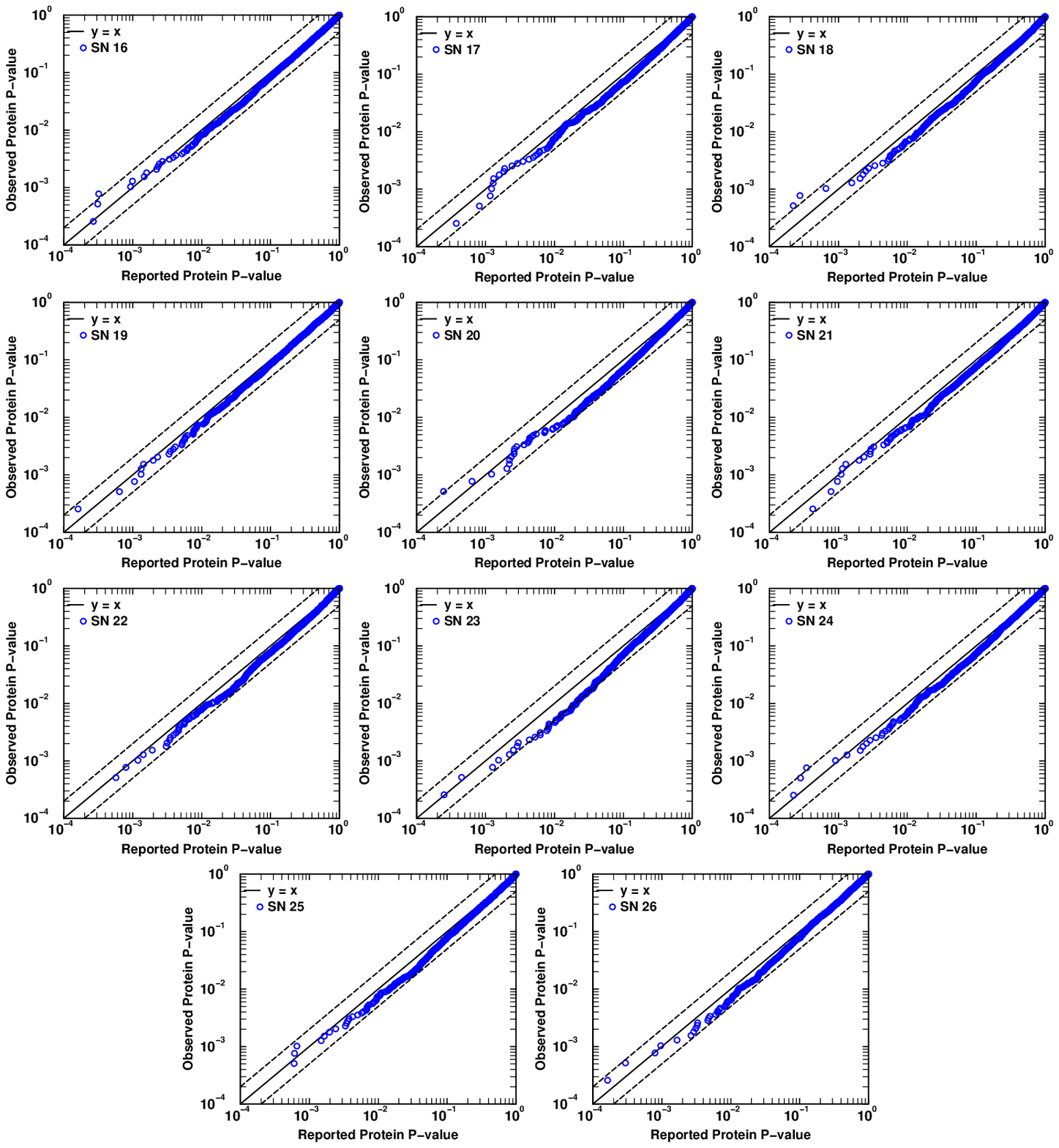}
\end{center}
\caption{{\bf Accuracy assessment of the protein {\it P}-value 
  using data group 2 and RAId\_aPS's peptide $E$-values
  when the selected scoring function is Hyperscore}. 
 All spectra in data group 2 are used to search the {\it Escherichia coli} database
  with precursor-ion mass tolerance $\pm$ 0.033 Da., product-ion mass tolerance 
  $\pm$ 0.033 Da. and up-to-2 missed cleavage sites allowed. The accuracy of the reported protein {\it P}-value  is evaluated by plotting it versus the observed protein {\it P}-value, see main text for details. The closer the above curves are to the $ y = x $ line the more accurate are the reported  protein {\it P}-values. The two dash lines, $y = 2x$ and $y = 0.5x$, are provided as visual guides regarding how close/off the reported protein {\it P}-values are to the $ y = x $ line.} 
\label{Fig:S8}
\end{figure*} 

\clearpage

\begin{figure*}[htbp]
\begin{center}
\includegraphics[width=0.92\textwidth,angle=0]{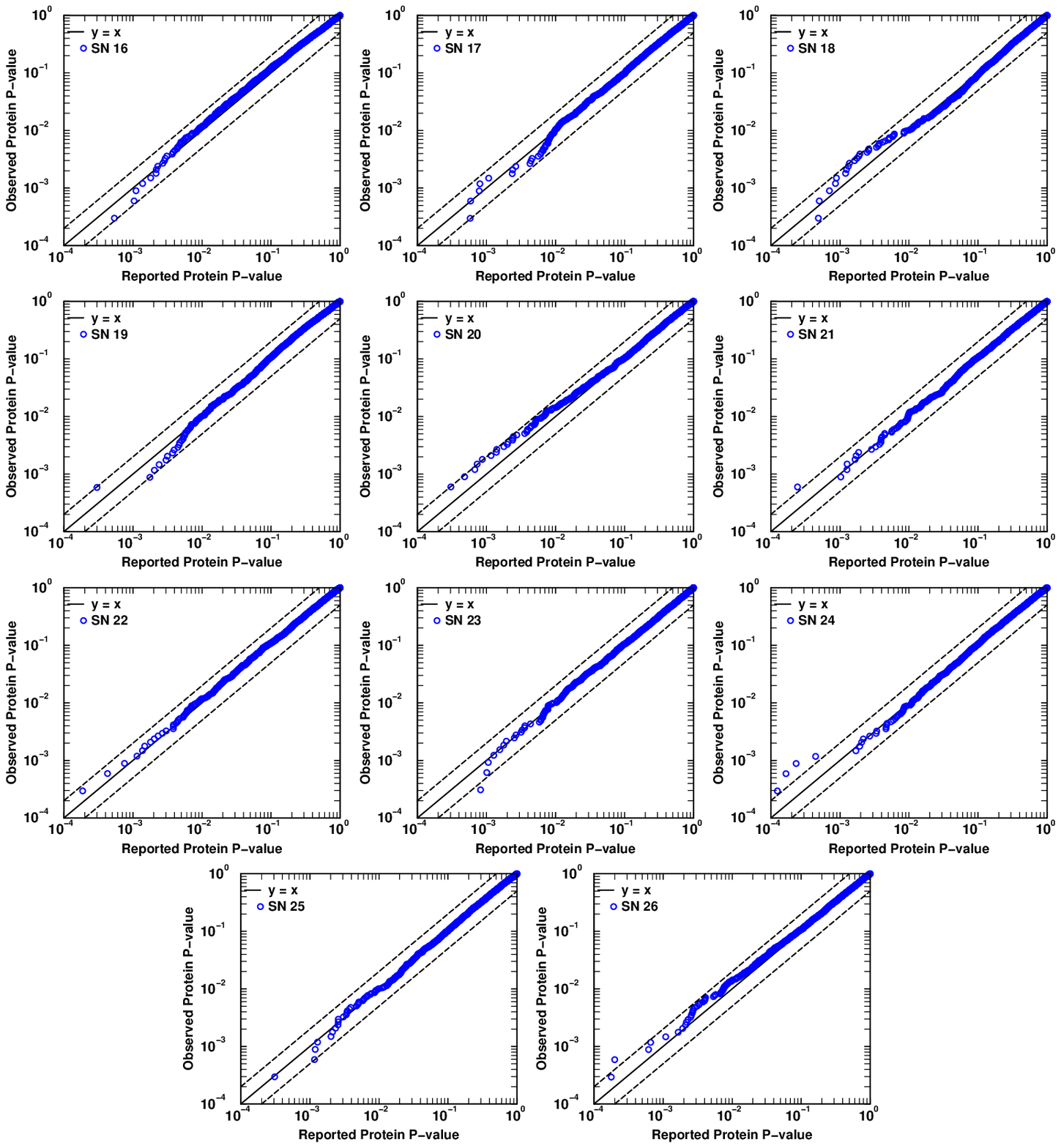}
\end{center}
\caption{ {\bf Accuracy assessment of the protein {\it P}-value 
  using data group 2 and RAId\_aPS's peptide $E$-values
  when the selected scoring function is Kscore}. 
 All spectra in data group 2 are used to search the {\it Escherichia coli} database
  with precursor-ion mass tolerance $\pm$ 0.033 Da., product-ion mass tolerance 
  $\pm$ 0.033 Da. and up-to-2 missed cleavage sites allowed. The accuracy of the reported protein {\it P}-value  is evaluated by plotting it versus the observed protein {\it P}-value, see main text for details. The closer the above curves are to the $ y = x $ line the more accurate are the reported  protein {\it P}-values. The two dash lines, $y = 2x$ and $y = 0.5x$, are provided as visual guides regarding how close/off the reported protein {\it P}-values are to the $ y = x $ line.} 
\label{Fig:S9}
\end{figure*} 

\clearpage

\begin{figure*}[htbp]
\begin{center}
\includegraphics[width=0.95\textwidth,angle=0]{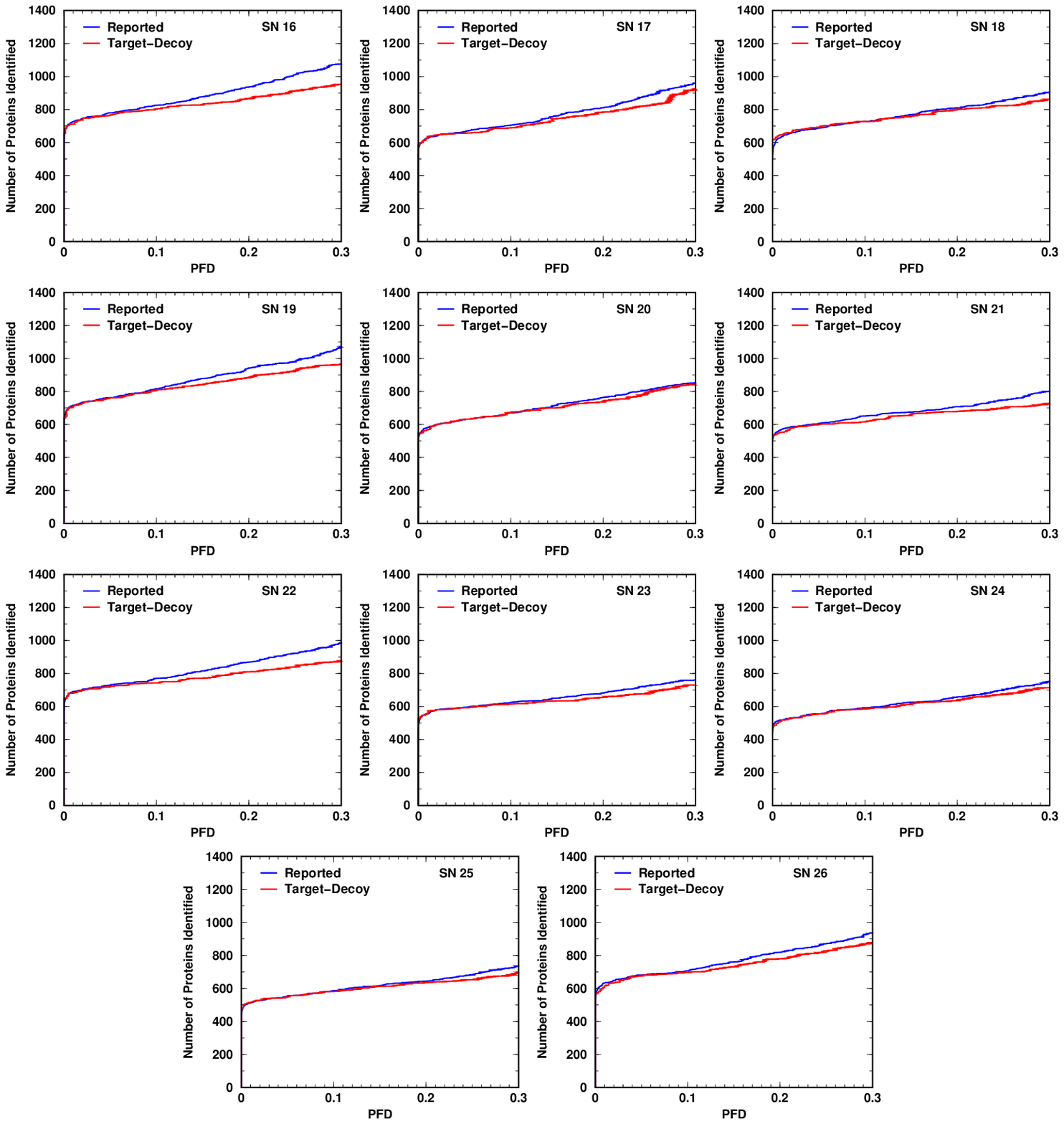}
\end{center}
\caption{  {\bf The agreement between the $E$-value based PFD and 
 the target-decoy based PFD when the peptide $E$-values are from 
 RAId\_aPS(XCorr)}. 
 All spectra in data group 2 (MS/MS spectra SN16-26) 
 are used to search the {\it Escherichia coli} database
  with precursor-ion mass tolerance $\pm$ 0.033 Da., product-ion mass tolerance 
  $\pm$ 0.033 Da. and up-to-2 missed cleavage sites allowed. The $E$-value based 
  PFD is computed by using the reported protein {\it E}-value in Sori\'c's formula, while the target-decoy based PFD is obtained from the ratio of the number of 
   identified decoy proteins to the total number of identified proteins (target + decoy) for a given {\it E}-value cutoff.  Within each panel, the closer 
  the two curves are to each other the better.} 
\label{Fig:S10}
\end{figure*} 

\clearpage

\begin{figure*}[htbp]
\begin{center}
\includegraphics[width=0.95\textwidth,angle=0]{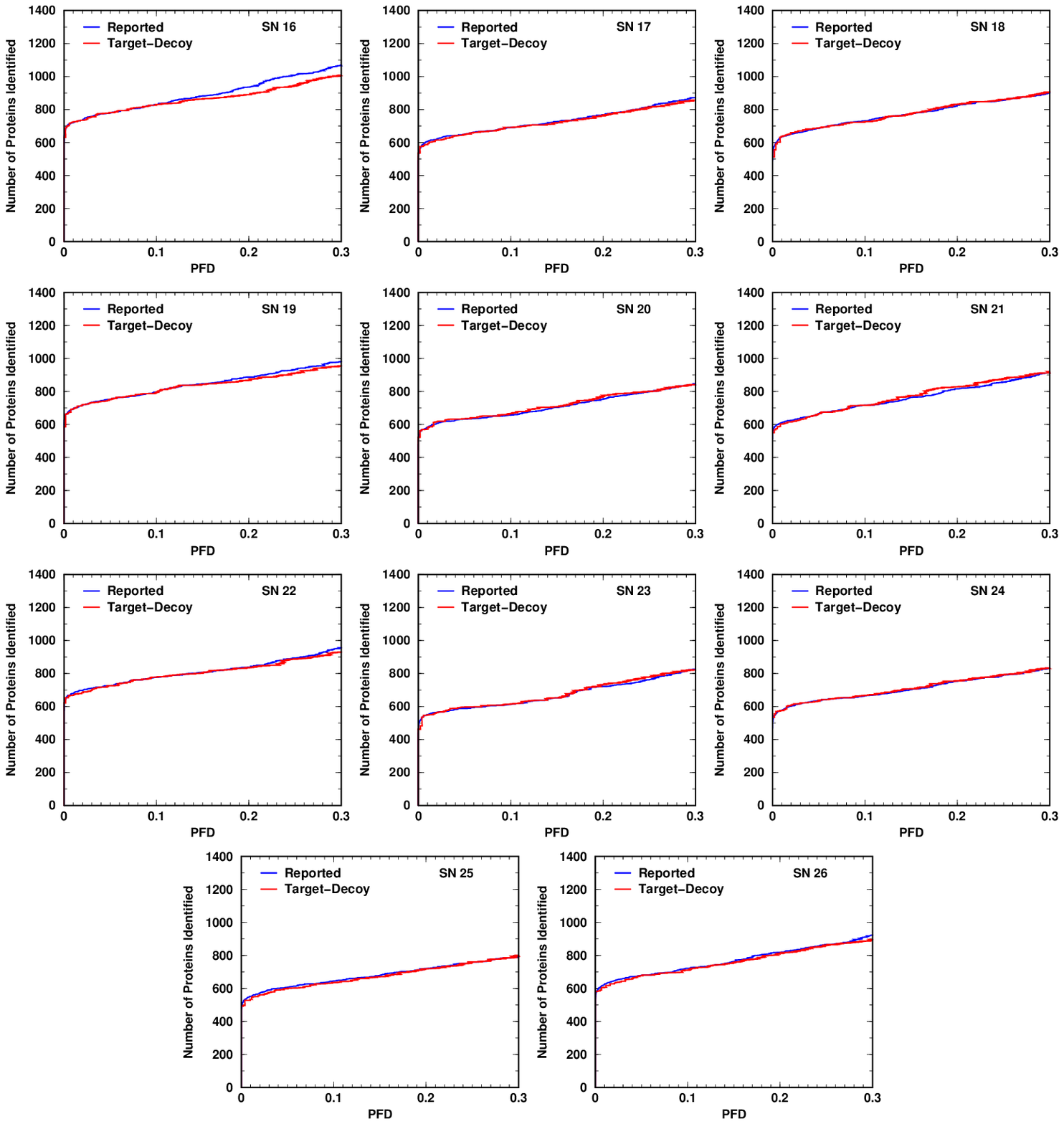}
\end{center}
\caption{ {\bf The agreement between the $E$-value based PFD and 
 the target-decoy based PFD when the peptide $E$-values are from 
 RAId\_aPS(Hyperscore)}. 
 All spectra in data group 2 (MS/MS spectra SN16-26) 
 are used to search the {\it Escherichia coli} database
  with precursor-ion mass tolerance $\pm$ 0.033 Da., product-ion mass tolerance 
  $\pm$ 0.033 Da. and up-to-2 missed cleavage sites allowed. The $E$-value based 
  PFD is computed by using the reported protein {\it E}-value in Sori\'c's formula, while the target-decoy based PFD is obtained from the ratio of the number of 
   identified decoy proteins to the total number of identified proteins (target + decoy) for a given {\it E}-value cutoff.  Within each panel, the closer 
  the two curves are to each other the better.} 
\label{Fig:S11}
\end{figure*} 

\clearpage

\begin{figure*}[htbp]
\begin{center}
\includegraphics[width=0.95\textwidth,angle=0]{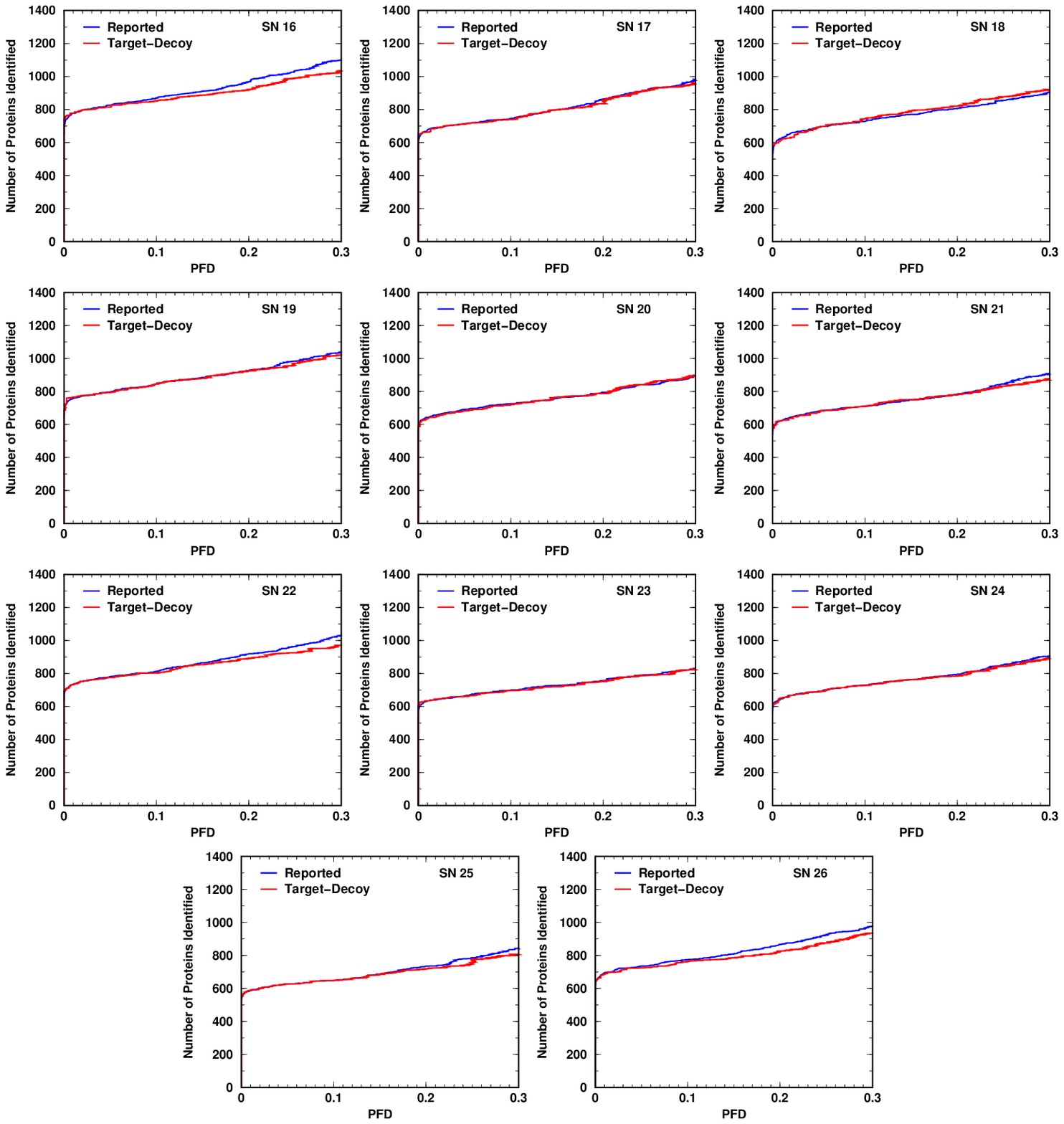}
\end{center}
\caption{ {\bf The agreement between the $E$-value based PFD and 
 the target-decoy based PFD when the peptide $E$-values are from 
 RAId\_aPS(Kscore)}. 
 All spectra in data group 2 (MS/MS spectra SN16-26) 
 are used to search the {\it Escherichia coli} database
  with precursor-ion mass tolerance $\pm$ 0.033 Da., product-ion mass tolerance 
  $\pm$ 0.033 Da. and up-to-2 missed cleavage sites allowed. The $E$-value based 
  PFD is computed by using the reported protein {\it E}-value in Sori\'c's formula, while the target-decoy based PFD is obtained from the ratio of the number of 
   identified decoy proteins to the total number of identified proteins (target + decoy) for a given {\it E}-value cutoff.  Within each panel, the closer 
  the two curves are to each other the better.} 
\label{Fig:S12}
\end{figure*}



\end{document}